\documentstyle[12pt,epsf]{article} 
\newcommand{\onefigure}[2]{\begin{figure}[htbp] 
         \begin{center}\leavevmode\epsfbox{#1}\end{center}\caption{#2}
         \end{figure}}

\advance\voffset by -1.8cm
\advance\hoffset by -1.9cm
\def\be{\begin{equation}}
\def\ee{\end{equation}}

\def\bbbz {{\sf Z\!\!Z}}

\def\bbbc{{\mathchoice {\setbox0=\hbox{$\displaystyle\rm C$}\hbox{\hbox
to0pt{\kern0.4\wd0\vrule height0.9\ht0\hss}\box0}}
{\setbox0=\hbox{$\textstyle\rm C$}\hbox{\hbox
to0pt{\kern0.4\wd0\vrule height0.9\ht0\hss}\box0}}
{\setbox0=\hbox{$\scriptstyle\rm C$}\hbox{\hbox
to0pt{\kern0.4\wd0\vrule height0.9\ht0\hss}\box0}}
{\setbox0=\hbox{$\scriptscriptstyle\rm C$}\hbox{\hbox
to0pt{\kern0.4\wd0\vrule height0.9\ht0\hss}\box0}}}}

\def\pmb#1{\setbox0=\hbox{#1}%
 \kern-.025em\copy0\kern-\wd0
 \kern.05em\copy0\kern-\wd0
 \kern-.025em\raise.0433em\box0 }
\def\sq{\hbox{\rlap{$\sqcap$}$\sqcup$}}
\def\qed{\ifmmode\sq\else{\unskip\nobreak\hfil
\penalty50\hskip1em\null\nobreak\hfil\sq
\parfillskip=0pt\finalhyphendemerits=0\endgraf}\fi}

\def\sqr#1#2{{\vcenter{\vbox{\hrule height.#2pt
                        \hbox{\vrule width.#2pt height#1pt \kern#1pt
                        \vrule width.#2pt}
                        \hrule height.#2pt}}}}

\newcommand{\bea}{\begin{eqnarray}}
\newcommand{\eea}{\end{eqnarray}}
\newcommand{\nn}{\nonumber}
\newcommand{\bean}{\begin{eqnarray*}}
\newcommand{\eean}{\end{eqnarray*}}

\newcommand{\myref}[1]{(\ref{#1})}
\newcommand{\unit}{1}
\newcommand{\mat}[4]{\left(\begin{array}{cc} #1 & #2 \\ #3 & #4 \end{array}\right)}
\newcommand{\vect}[2]{({#1\atop #2})}
\begin{document}

\thispagestyle{empty}
\def\thefootnote{\fnsymbol{footnote}}
\begin{flushright}
  DAMTP-98-6 \\
  MIT-CTP-2712 \\
  hep-th/9801205
\end{flushright} \vskip 0.5cm

\begin{center}\LARGE
{\bf Open string -- string junction transitions}
\end{center} \vskip 0.8cm
\begin{center}\large
       Matthias R.\ Gaberdiel%
       \footnote{E-mail  address: {\tt M.R.Gaberdiel@damtp.cam.ac.uk}}%
       \end{center}
\begin{center}
Department of Applied Mathematics and Theoretical Physics    \\
University of Cambridge \\
Silver Street, 
Cambridge CB3 9EW,  
England
\end{center}
\vskip0.3cm
\begin{center}\large
        Tam\'as Hauer and Barton Zwiebach%
       \footnote{E-mail  addresses: {\tt 
hauer@mit.edu, zwiebach@irene.mit.edu}}%
       
       \end{center}
\vskip0.3cm
\begin{center}
Center for Theoretical Physics\\
LNS and Department of Physics, MIT\\
Cambridge, MA 02139, USA
\end{center}
\vskip 1em
\begin{center}
January 1998
\end{center}
\vskip 1cm
\begin{abstract}

It is confirmed that geodesic string junctions are necessary to
describe the gauge vectors of symmetry groups that arise in the
context of IIB superstrings compactified in the presence of nonlocal
7-branes. By examining the moduli space of 7-brane backgrounds for
which the dilaton and axion fields are constant, we are able to
describe explicitly and geometrically how open string geodesics can
fail to be smooth, and how geodesic string junctions then become 
the relevant BPS representatives of the gauge bosons. The mechanisms
that guarantee the existence and uniqueness of the BPS representative
of any gauge vector are also shown to generalize to the case where the
dilaton and axion fields are not constant.

\end{abstract}

\vfill

\setcounter{footnote}{0}
\def\thefootnote{\arabic{footnote}}
\newpage

\renewcommand{\theequation}{\thesection.\arabic{equation}}

\section{Introduction}
\setcounter{equation}{0}

Type IIB superstring theory possesses BPS states $({p\atop q})$ 
which are bound states of $p$ fundamental and $q$ D-strings
\cite{wittenbound}, where $p$ and $q$ are relatively prime and
represent the charges of the resulting string under the  NS-NS and 
RR antisymmetric tensors of the theory, respectively. The theory also
possesses a corresponding set of $[p,q]$ 7-branes.   
An elementary string, {\it i.e.~}a $({1\atop 0})$-string can end on a
D-string {\it i.e.~}a $({0\atop 1})$-string, and in this process part
of the D-string turns into a $({1\atop 1})$-string in order to satisfy
the charge conservation. This  configuration is  more appropriately 
thought of as a junction where three different strings join
\cite{aharony,schwarzreview}. The relevant conservation laws were
explained by Schwarz who conjectured that string junctions would 
represent physically relevant BPS states \cite{schwarzreview}. 

In~\cite{GZ} we showed that there exists an interesting physical
configuration where these string junctions seem to play an important
role. This configuration is described by the compactification of type
IIB on a two sphere in the presence of twenty-four 7-branes, not all of
which are of the same $[p,q]$ type; this corresponds in F-theory to
the compactification on an elliptically fibered K3 \cite{vafa}. As the
K3 develops an ADE-singularity, gauge bosons in the corresponding
group become massless \cite{bikmsv}. From the point of view of type
IIB, these BPS states may be thought of as  certain open 
strings that begin and end on various 7-branes \cite{johansen}. 
However, as argued in \cite{GZ}, open strings do not seem to suffice,
and open string junctions seem necessary depending on the positions 
of the relevant 7-branes. Moreover (and this was the main motivation
in \cite{GZ}), the pronged string states manifestly carry the correct
gauge charges and this enables one to reproduce the multiplicative
structure of the exceptional groups. This leads to a compelling brane
picture for the gauge enhancement of exceptional symmetry.  

Subsequently, several works have appeared giving detailed and
convincing arguments that string junctions preserve supersymmetry
\cite{AHK,junctions,sennetwork,rey,krogh,matsuo}. A second physical
application for string junctions has also been proposed recently by
Bergman \cite{OB}, who gave evidence that three string junctions
should represent one-quarter-BPS states in a string picture of $N=4$,
four-dimensional supersymmetric gauge theory with gauge groups of rank
bigger than one.  Type IIB theory has also $[p,q]$ five-branes and
their junctions are relevant to five-dimensional gauge theory
\cite{AHK}. While the physics of five-branes  is rather different 
from that of strings, the mathematical methods used to study
five-brane junctions have relevance to string junctions.
\smallskip

In this paper we return to the subject of \cite{GZ}. The 7-branes in
question appear as points on the two-sphere and extend in the eight
non-compact directions, and the background of the 7-branes induces a
nontrivial metric on the two-sphere. Open strings that join the
7-branes and represent BPS states of the 8-dimensional field theory are
naturally realized as smooth geodesics. A preliminary study of the
metric on $S^2$ suggested that it was unlikely that one could have
smooth geodesics in all requisite homotopy classes \cite{GZ}, and it
was shown that open strings could become string junctions as the
string crosses a 7-brane. In this process an extra prong is created,
and this is U-dual to the Hanany-Witten effect
\cite{hananywitten}. This suggested a picture where a given BPS-state 
could be represented sometimes by a smooth geodesic string, and
sometimes by a geodesic junction, where all prongs are smooth.
The physical consistency of this  proposal hinges on three
properties: 
\begin{list}{(\roman{enumi})}{\usecounter{enumi}}
\item {\it Necessity of junctions.} As we change the positions of the
7-branes, the open string geodesic representing a gauge vector can
fail to be smooth, and thus fail to be a valid representative. 
\item {\it Existence.} In the situation of (i), a geodesic string
junction of mass lower than that of the broken open string exists and
represents the gauge vector. 
\item {\it Uniqueness.} For any configuration of 7-branes the
minimal mass object representing a gauge vector is unique; it is 
{\it either} a smooth geodesic open string {\it or} a geodesic string
junction. 
\end{list}
The main purpose of the present paper is to explain the mechanisms
that guarantee that (i), (ii) and (iii) hold. 
This will also provide an indirect (and explicit) confirmation of the 
Hanany-Witten effect. We shall mainly consider particularly simple
situations where the transitions can be understood very explicitly,
and we shall demonstrate how the arguments generalize. Along the way
we shall also learn much about the metric on the two-sphere. 
\smallskip

The IIB backgrounds we are considering can also be viewed as an
M-theory compactification on an elliptic K3 in the limit as the
elliptic fibers, the tori, are of vanishing area
\cite{mtheory,schwarzreview,AHK,krogh}. In this framework the relevant
BPS states arise from membranes that wrap around supersymmetric cycles
of genus zero, and in the limit of zero area, such cycles are expected
to project to geodesics on the two-sphere base of the elliptic 
fibration. This viewpoint gives strong indirect evidence for the
validity of (i)--(iii). Indeed, open strings and string junctions
related by crossing operations should define K3 cycles belonging to
the same homology class, and in a K3 cycles of genus zero in any given
homology class have unique supersymmetric representatives. Moreover,
we should expect the supersymmetric representatives to vary
continuously as we vary the moduli of the K3. While this line of
argumentation, familiar to experts, could conceivably be developed
into a proof of (i)--(iii), we shall not attempt this here. We shall
rather work directly with the IIB picture, and analyze the two-sphere
with its nontrivial metric and 7-branes in detail. This picture has
its merits: it is explicit, BPS states are easily visualized, and 
the role of string junctions is clearly exhibited.

In order to examine the various string geodesics and string
junctions, it is often necessary to separate the different 7-branes
(that define the singularity when they coincide), and we shall refer
to this as resolving the singularity. When the branes are separated,
however, the metric on the two-sphere is typically a rather
complicated function of $\tau=a+ie^{-\phi}$ (where $a$ is the axion
and $\phi$ the dilaton), which in itself is a nontrivial multivalued
function on the sphere. We show that partial resolutions of the above
singularities are possible while maintaining a constant value of
$\tau$ throughout the sphere. For example the $so(8)$ singularity ($D_4$) 
defined by  six coincident 7-branes can be resolved at constant $\tau=
\exp (i\pi/3)$ into three separate singularities, each containing two
coincident 7-branes. Analogous resolutions are also possible for the
$E_6$, $E_7$ and $E_8$ singularities.  

These partial resolutions fit nicely within the moduli space of
compactifications at constant $\tau$ which consists of three branches:
the branch where $\tau$ can take an arbitrary (constant) value which
was described by Sen \cite{senorientifold}, and the two branches 
described by Dasgupta and Mukhi \cite{dasgupta} that are consistent
with particular values of constant $\tau$. For the former a
description in terms of 7-branes was already given by Sen, and for the
latter we shall provide such an explicit description in this
paper. For example, on one of the two additional branches, the generic
situation is described by twelve identical singularities, each
defined by two mutually non-local 7-branes, and we show how collisions
of these singularities generate all enhanced symmetry points on this
branch. The main tool in this discussion is the understanding of how
7-branes are transformed as 7-brane {\it branch cuts} are moved across
other 7-branes. This enables us to relate singularities defined by two
types of 7-branes to singularities defined by three types of
7-branes; the latter are the conventional presentations for
exceptional singularities presented by Johansen \cite{johansen}.
\medskip

For the discussion of the transition between open string geodesics and
string junctions, we shall mainly consider partial resolutions (at
constant $\tau$) into three singularities. At constant $\tau$, 
all $({p\atop q})$ strings feel the same metric (up to a constant
factor that is due to the tension), and there exists a coordinate $w$,
in which all $({p\atop q})$ geodesics are straight lines. Moreover,
the metric singularities are of conical type, and they can thus be
represented (in the $w$-plane) by excising a wedge and identifying the
edges of the wedge, with the proviso that strings change their $(p,q)$
character as they cross the seam. We can then examine the fate of
a direct geodesic (between two of the singularities $P_1$ and $P_2$,
say) as the position of the third singularity $Q$ changes. It will
become apparent that the direct geodesic ceases to be smooth as the
straight line between $P_1$ and $P_2$ crosses the wedge (see (i)
above). In this case, as we shall demonstrate, a string junction is
created whose overall mass is precisely described by the direct
distance between $P_1$ and $P_2$, and this shows directly that the
three string junction has strictly lower length than
that of the broken direct geodesic ((ii) above) and that both
representatives cannot simultaneously be smooth ((iii) above). Given
that a three singularity resolution already exists for $so(8)$, this
implies that string/junction transitions are not a phenomenon peculiar
to exceptional groups; rather they are a generic non-perturbative
phenomenon. 

The above discussion can be generalized to backgrounds where 
$\tau$ is not necessarily constant. In particular, we can still find a
simple expression for the overall length of the string junction as the
distance between two points, and the main conclusions are therefore
unaltered. We shall also show explicitly how for the $so(8)$
singularity, some of the indirect geodesics fail to be smooth as the
singularity (that corresponds to the orientifold plane in the
corresponding type I description) is resolved.
\medskip

This paper is organized as follows. In section 2 the various 
conventions are explained in detail, and it is shown how the
characterization of branes changes as branch cuts are moved across
other branes. In section 3 we give a brane description of the moduli
space of the constant $\tau$ compactifications, and we show how to
recover the $D_4$ and $E_6, E_7, E_8$ singularities by collision of 
simpler singularities. In section 4 we describe explicitly the
transitions between direct strings, string junctions and indirect
strings for partial resolutions compatible with constant $\tau$. In
section 5 we extend these results to non-constant $\tau$
backgrounds. In section 6 we show, in the context of a familiar  
non-constant $\tau$ background, the absence of smooth indirect
geodesics for a certain class of strings. Finally, in section 7 
we offer some concluding remarks
and discuss open questions.

\section{The description of the covering space}
\setcounter{equation}{0}

The main object of our discussion is type IIB string theory
compactified on a two-sphere in the presence of twenty-four
parallel 7-branes which extend along the eight uncompactified
directions. In this section we shall explain our various conventions
for the monodromies of the 7-branes (not all of which may be mutually
local) on the compactifying sphere.  
In particular, we shall explain how to introduce branch cuts on the
two-sphere, and how the description changes as branch cuts are moved 
across branes.

\subsection{Seven-branes, monodromies and $(p,q)$-strings}  

Let us denote by $\tau=a+i e^{-\phi}$ the complex combination of the
dilaton field $\phi$ and the axion field $a$ in type IIB. In the
presence of 7-branes, $\tau$ is not a single-valued function on the
sphere, and the way in which it fails to be single-valued
characterizes the 7-branes that are located on the
sphere.\footnote{The following clarifies (and corrects) the
conventions that were used in \cite{GZ}.} In order to describe this in
some detail, let us denote by $z_i$ ($i=1,\ldots,24$) the locations of
the 7-branes on the sphere, and let $S_0$ be the punctured sphere
where all twenty-four points $z_i$ have been removed. We pick a
reference point $z_0\ne z_i$ on the (punctured) sphere, and we
introduce a basis of generators $\gamma_i$ for the homotopy group
$\pi_1(S_0)$ of $S_0$ relative to $z_0$. (Here $\gamma_i(t)$ is a path
on $S_0$ that starts at $\gamma_i(0)=z_0$, encircles $z_i$
anti-clockwise, and ends again at $\gamma_i(1)=z_0$.)  

For the configuration we have in mind, the monodromy of $\tau$ can be
described by a collection of matrices $M_i\in\mbox{SL}(2,\bbbz)$,
where 
\be
\label{monodromy}
\tau(\gamma_i(1)) = M_i \tau(\gamma_i(0)) \,,
\ee
and $M\in\mbox{SL}(2,\bbbz)$ acts on $\tau$ in the usual way
\be
M \tau = {a \tau + b \over c \tau + d} \,, \qquad \mbox{where} \qquad
M=\pmatrix{a & b\cr c & d} \,.
\ee
As $\tau$ is single-valued in every simply-connected subset of $S_0$,
it follows from (\ref{monodromy}) that the $M_i$ define a
group homomorphism $M:\pi_1(S_0)\rightarrow \mbox{SL}(2,\bbbz)$. 

In order to represent a fundamental 7-brane (rather than a bound state
of different 7-branes),  $M_i$ must be of the form  
\be
\label{genmon}
M_{p,q} = g_{p,q}\,  T\,  g_{p,q}^{-1} = 
\pmatrix{1-pq& p^2\cr -q^2& 1+pq}\,,
\ee
where
\be
T = \pmatrix{1 & 1 \cr 0 & 1} \qquad \hbox{and} \qquad
g_{p,q} = \pmatrix{p & r \cr q & s} \,,\qquad ps -qr =1\,.
\ee
In particular, the different monodromy matrices are thus
$\mbox{SL}(2,\bbbz)$ conjugates of $T$, the monodromy matrix of the
$[1,0]$ D7-brane \cite{douglasli}.  
With respect to the chosen generators  
of the homotopy group of $S_0$, we say that the 7-brane at $z_i$ is of
type $[p,q]$ if $M_i=M_{p,q}$. This defines the $[p,q]$ label of a
given 7-brane up to an overall sign, as $M_{-p,-q}=M_{p,q}$. We should
stress that this identification depends on the choice of $z_0$ and
$\gamma_i$.  

The system possesses a global $\mbox{SL}(2,\bbbz)$ symmetry: for
$g\in\mbox{SL}(2,\bbbz)$, we can replace 
\be
\label{sl2ztrans}
\tau \mapsto g \tau\,, \qquad \hbox{and then} \qquad
M_i \mapsto g M_i g^{-1} \,,
\ee
since $g \tau(\gamma_i(1))=g M_i \tau(\gamma_i(0))=
(g M_i g^{-1}) \, g \tau(\gamma_i(0))$.
\smallskip

Given the multi-valued function $\tau$ on $S_0$, we can define a
metric on $S_0$ by \cite{greeneetal}  
\be
ds^2 = \tau_2  \, \eta(\tau)^2 \bar\eta (\bar\tau)^2 \prod_i
(z-z_i)^{-1/12} (\bar z - \bar z_i)^{-1/12} \, dz d\bar z\,,
\ee
where $\tau_2=\mbox{Im}(\tau)$, and
\be
\label{expeta}
\eta^2 (\tau) = q^{1/12} \prod_{n=1}^\infty (1-q^n)^2\,, \quad 
q=\exp (2\pi i\tau)\,, 
\ee
is the square of the Dedekind eta-function which satisfies
\be
\label{etatr}
\eta^2 (-1/\tau) = -i\tau \eta^2 (\tau) \,, \quad
\eta^2 (\tau+1) = e^{i\pi/6} \, \eta^2(\tau)\,.
\ee
It is straightforward to show that this metric is invariant under the
global $\mbox{SL}(2,\bbbz)$ action on $\tau$, and this implies that
the metric is single-valued on $S_0$. 
\smallskip

Type IIB theory possesses different strings which are labeled by
$({p\atop q})$, where $p$ and $q$ are coprime integers. The masses of
the states associated to $({p\atop q})$ strings must also take into
account the string tension $T_{p,q}$ of the $({p\atop q})$ string
\cite{schwarz}, 
\be
T_{p,q} = {1\over \sqrt{\tau_2}}\, | p-q\tau \,|\,.
\ee
It is then natural to introduce the length element
$ds_{p,q}=T_{p,q}ds$ which measures correctly the mass of the
corresponding string, and the corresponding effective metric
$ds^2_{p,q}$ 
\be
\label{effmet}
ds_{p,q} = \left| h_{p,q}(z) \, dz \right| \,, 
\quad\hbox{with}\quad h_{p,q}(z) = (p-q\tau)\,
\eta^2(\tau) \prod_i (z-z_i)^{-1/12}\,.
\ee
For every fixed label $({p\atop q})$, $ds_{p,q}^2$ describes a
continuous metric on a certain covering space $\widetilde{S}_{p,q}$ of
$S_0$. If we define the action of $\mbox{SL}(2,\bbbz)$ on 
$({p\atop q})$ by 
\be
\label{invariance}
\pmatrix{p\cr q} \mapsto \pmatrix{p'\cr q'}
=\pmatrix{a& b\cr c& d}\cdot \pmatrix{p\cr q}\,, \qquad \mbox{where}
\qquad g=\pmatrix{a& b\cr c& d}\in\mbox{SL}(2,\bbbz) \,,
\ee
then for $\tau'=g\tau$, we have $T_{p,q}(\tau) = T_{p',q'}(\tau')$. 
This implies that up to a global transformation of $\tau$ by an
element in $\mbox{SL}(2,\bbbz)$, the spaces $\widetilde{S}_{p,q}$ and
$\widetilde{S}_{p',q'}$ are the same.

\subsection{Introducing branch cuts}

Different strings can end on different 7-branes, but there is no
$\tau$-independent description of which string can end on which
7-brane. In order to discuss this issue it is therefore necessary to
introduce branch cuts on $S_0$ and to choose a fixed branch $\tau_0$
of $\tau$ on the space $S_c=S_0 - \{\mbox{branch cuts}\}$. (The branch
cuts are chosen so that $S_c$ is simply connected.) 

There is a large amount of freedom how to choose the branch cuts (and
we shall later take some advantage of this fact) but there is a
particularly simple choice: for each $z_i$ we introduce a branch
cut $C_i$ from $z_i$ to $z_0$ so that the path $\gamma_i$ does not
cross $C_i$ (and that it only touches the branch cut $C_i$ at the
common end point $z_0$).\footnote{The different choices for $z_0$ and
$\gamma_i$, and the different possible branch cuts $C_i$ are in
one-to-one correspondence: every choice of $z_0$ and $\gamma_i$
determines a family of cuts $C_i$ by the above description, and
conversely, every such family of cuts determines $z_0$ (the common
endpoint of the $C_i$) and $\gamma_i$ up to homotopy.} It then follows
that as $\tau$ crosses $C_i$ anti-clockwise, $\tau$ jumps to $K_i\tau$
where  
\be
\label{relation}
K_i = M_i^{-1} \,.
\ee
We should stress that this relation only holds for this particular
choice of cuts, and that for a different choice of cuts,
(\ref{relation}) has to be replaced by $K_i = D_i M_i^{-1} D_i^{-1}$,
where $D_i$ is some matrix in $\mbox{SL}(2,\bbbz)$; this will be
explained in more detail in the next subsection. For this choice of
cuts we can then say that a 7-brane is of type $[p,q]$ if
$K_i=M_{p,q}^{-1}$.  
\smallskip

Given $({p\atop q})$ and $\tau_0$ on $S_c$, we can define a metric
on $S_c$ by formula (\ref{effmet}). This metric is not continuous
across the branch cut, and in order for it to become continuous, we 
have to choose the convention that the labels of the string 
$({p \atop q})$ change to $K_i ({p \atop q})$ as we cross the branch
cut anti-clockwise; the metric is then continuous because of 
(\ref{invariance}). As $({p\atop q})$ is locally constant, it follows
that the $K_i$ induce a group homomorphism
$K:\pi_1(S_0)\to\mbox{SL}(2,\bbbz)$; this can be easily understood
from Fig.~\ref{fig1}. In this example the monodromy corresponding to
the curve $\gamma_1\circ\gamma_2\circ\gamma_3$ is $M_1M_2M_3=\unit$
(see figure); on the other hand, following the curve $\gamma$, one
crosses first the cut $C_1$, then $C_2$ and finally $C_3$ which
corresponds to 
\bean
K_3K_2K_1 = M_3^{-1}M_2^{-1}M_1^{-1} = (M_1M_2M_3)^{-1} = \unit.
\eean
\onefigure{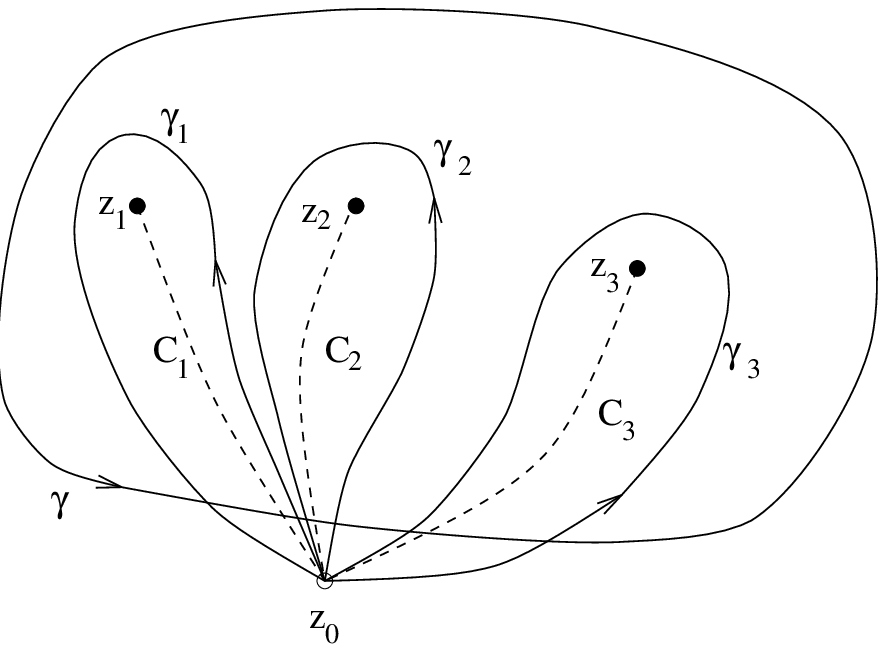}{The branes can be characterized by either the
monodromies or the discontinuity at their branch cuts. The matrices
$M$ and $K$ both induce a group homomorphism
$\pi_1(S_0)\to\mbox{SL}(2,\bbbz)$. \label{fig1}}

We can thus summarize our convention as (see Fig.~\ref{fig0}) 
\begin{verse}
\noindent {\it Upon anticlockwise crossing of the branch cut of a 7-brane
with matrix $K$, $\tau$ transforms to $\tau\to K\tau$ and 
$({p \atop q})$ transforms to $({p \atop q})\to K ({p \atop q})$.}
\end{verse}
\onefigure{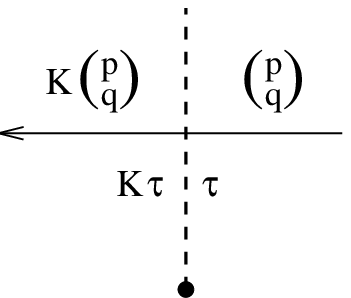}{As the  branch cut is crossed 
     anticlockwise both
     $\tau$  and  $\vect{p}{q}$ are transformed by $K$. \label{fig0}}

We shall now say that a $({p \atop q})$ string can end on a brane
described by $K$ if and only if $K=M_{p,q}^{-1}$. If $K$ is defined
with respect to the specific choice of cuts described above, then this
rule amounts to saying that a $({p \atop q})$-string can only end on a
$[p,q]$ brane. 
\smallskip

In this paper we shall mainly deal with three types of branes 
${\bf A}$, ${\bf B}$ and ${\bf C}$, whose corresponding $[p,q]$ labels
(with respect to some choice of $z_0$ and $\gamma_i$) are $[1,0]$,
$[1,-1]$ and $[1,1]$ respectively. Across the corresponding cuts
$C_i$, the labels $({p \atop q})$ and $\tau$ then change according to  
\bea
\label{expl}
{\bf A}=& [1,0]: \,\,K_A = & M_{1,0}^{-1} = T^{-1} = 
\pmatrix{1 & -1 \cr 0 & 1} \,, \nonumber\\ 
{\bf B}=& [1,-1]: K_B = & M_{1,-1}^{-1} =  S T^{2} 
             = \pmatrix{0 & -1 \cr 1 & 2} \,, \\
{\bf C}=& [1,1]: \,\, K_C = & M_{1,1}^{-1} =  T^{2}S 
             = \pmatrix{2 & -1 \cr 1 & 0} \,,\nonumber
\eea  
where $S$ is the matrix  
$$
S = \pmatrix{ 0 & - 1 \cr 1 & 0 } \,.
$$

We shall adopt the convention that for a given arrangement of the
branes and their cuts, we shall write the corresponding matrices $K_i$
in the order (from right to left) in which we would cross the branch
cuts as we encircle the relevant branes in a large anticlockwise circle, 
and we shall use the same convention also for the labels ${\bf A}$,
${\bf B}$ and ${\bf C}$. For example, the list of branes
\be
\label{alist}
{\bf A_1 A_2} \cdots {\bf A_n}
\ee
represents a situation where the branes 
lie below the real axis, and all the branch cuts go {\it upwards}
along the imaginary axis after  crossing the real axis at points
$x_{A_1}<x_{A_2}<\cdots<x_{A_n}$.\footnote{In this configuration, 
$z_0=\infty$.} The corresponding $K$-matrices then
read $K_{A_1} K_{A_2} \cdots K_{A_n}$.

\subsection{Moving branch cuts across branes} 

Let us now explain how the $K$-matrices of the branes change as we
change the location of the cuts. First of all, it is clear that the
$K_i$ are locally independent of the choice of the cuts, and that the
matrix $K_i$ associated to the {\bf i}-brane only changes as the cut of
another 7-brane is moved through the position $z_i$ of this 7-brane.

For definiteness, let us consider the case where we move the cut $C_j$
of the  ${\bf j}$-brane clockwise through the position $z_i$ of an
${\bf i}$-brane. This operation can be implemented by performing a 
physically immaterial {\it local} $\mbox{SL}(2,\bbbz)$ transformation
where we let $\tau\to K_j \tau$ in the region enclosed by $C_j$ and
$C_j'$. (Indeed, this transformation makes $\tau$ continuous across
$C_j$ and creates a new discontinuity across $C_j'$.)

Let us consider the situation where the cut $C_i$ that emerges from
$z_i$ is inside the region that is defined by $C_j$ and $C_j'$.
Before we moved the cut, $\tau$ transformed to $K_i \tau$ as we
crossed $C_i$ anti-clockwise, and after we have moved $C_j$ to $C_j'$,
we get $K_j\tau\to K_jK_i \tau = K_j K_i K_j^{-1} (K_j\tau)$, 
implying that the  $K_i\to K_i'\equiv K_j K_i K_j^{-1}$ as a
consequence of the (clockwise) motion of $C_j$ through the 
{\bf i}-brane; this is illustrated in Fig.~\ref{fig2}. 
The shaded region denotes the region defined by the old and new
cuts. 
\onefigure{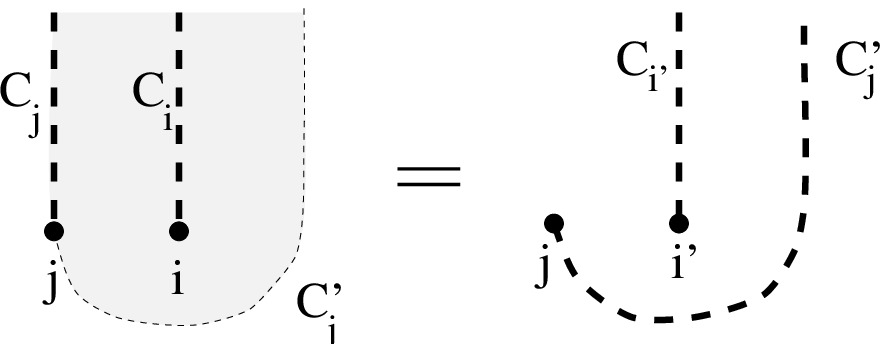}{The local $\mbox{SL}(2,\bbbz)$-transformation
     induced by $K_j$ in the shaded area relocates 
     the cut of the ${\bf j}$-brane and  changes the ${\bf
     i}$-brane to an ${\bf i'}$-brane. In the standard 
     presentation the right picture reads ${\bf i'j}$ as
     $C_j'$ is to the right of $C_i$ and thus we have 
     ${\bf ji}\longrightarrow{\bf i'j}$. 
     \label{fig2}}

The whole effect can be described as 
\be
\label{trcr} 
{\bf j~i } \to {\bf i'\, j}\,, \quad\hbox{with}\quad K_i'= 
K_j K_i K_j^{-1}\,.
\ee
This transformation respects, as expected, the product of the
$K$-matrices that is associated with the transformation of $\tau$
along a large circle surrounding the pair of branes 
$$
K_j K_i = \left( K_j K_i K_j^{-1} \right) K_j = K_i' K_j \,.
$$
\onefigure{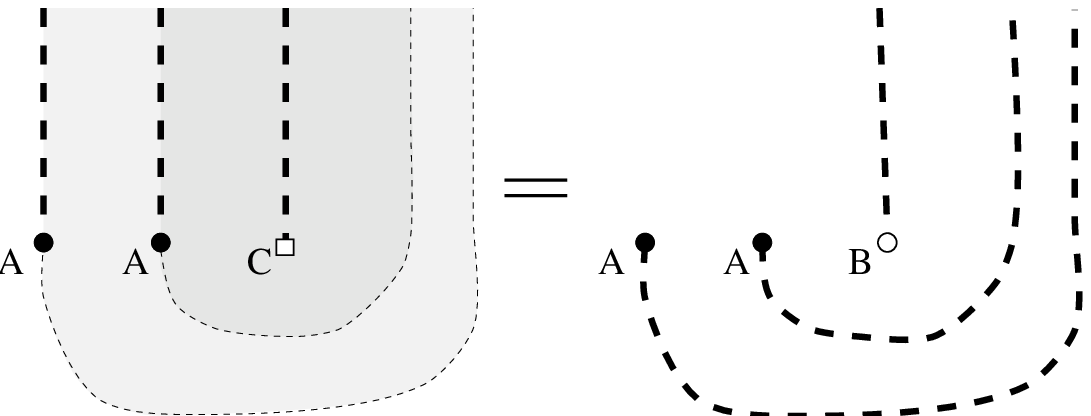}{The effect of a local
     $\mbox{SL}(2,\bbbz)$-transformation induced by $K_A$ in the
     light and by $K_A^2$ in the dark shaded area results in the
     relocation of both {\bf A}-cuts and the change of the {\bf
     C}-brane into a {\bf B}-brane: {\bf AAC}$\longrightarrow$ {\bf
     BAA}. \label{fig3}}

As an illustration of this, let us consider the 
configuration {\bf AAC} depicted in Fig.~\ref{fig3}\footnote{In
the figures of the present paper we denote the {\bf A}-branes by
filled circles, the {\bf B}-branes by empty circles and the 
{\bf C}-branes by empty squares}, and move successively
the cuts of the {\bf A} branes across the {\bf C} brane. In this
process we get $K_C \to K_A^2 K_C K_A^{-2} = K_B$, as is readily
verified using \myref{expl}. We can therefore write
\be
\label{ftrans}
{\bf AAC} \to {\bf BAA}\,.
\ee 
Similarly we have
\be
\label{strans}
{\bf ACB} \to {\bf CA'B} \to {\bf CBA} \,,
\ee
where $K_{A'} = K_C^{-1} K_A K_C$ as follows from moving the cut of 
{\bf C} anti-clockwise across {\bf A}. The final step arises
by moving the cut of {\bf B} anti-clockwise
across {\bf A}$'$ on account of $K_B^{-1}K_{A'} K_B= K_A$. We thus learn
that the {\bf A} brane ``commutes" with the {\bf BC} sequence.
\smallskip

In general, given a basepoint $z_0$ and contours
$\gamma_i$, we can first choose the cuts $C_i$ as
described in the previous subsection so that $K_i=M_i^{-1}$.
We can then obtain any
other choice of cuts by successively moving the cuts through the
singularities. It is then clear that the matrices $K_i$ that are
associated to a general family of cuts are of the form 
$K_i = D_i M_i^{-1} D_i^{-1}$ where $D_i \in \mbox{SL}(2,\bbbz)$.
\smallskip

It is now also clear that the prescription that a $({p \atop q})$
string can end on a brane with $K=M_{p,q}^{-1}$ is independent of
the choice of the cuts. As we move the cut corresponding to $K_j$
clockwise through the brane, $K$ becomes 
$K'=K_j K K_j^{-1} = M_{p',q'}^{-1}$, where 
$({p'\atop q'}) = K_j ({p \atop q})$. On the other hand, 
$({p \atop q})$ becomes $({p'\atop q'})$, and it follows that the
prescription is invariant as we change the cuts.

\section{Resolutions, Singularities and the Moduli space of
constant $\tau$}
\label{sec:modspace}
\setcounter{equation}{0}

The dynamics of $(p,q)$-strings in a general axion-dilaton background
is quite nontrivial. Although the metric is locally flat, the
coordinate in which geodesics are straight lines is different for
mutually nonlocal strings, and the transition functions between these
coordinates are rather complicated. In section~\ref{sec:nonconstant}
we shall examine the general case but before doing so we want to analyze
the problem for the special branches of the moduli space where $\tau$
is constant over $S^2$. As we shall see, all the important features of
the problem will already be visible in this case.

The subset of the moduli space where $\tau$ is constant has been
analyzed before by Sen \cite{senorientifold}, who considered only the
case of the $D_4$ singularity, and by Dasgupta and Mukhi
\cite{dasgupta}, who showed that there are two additional branches. 
We will briefly review their work, and then give brane descriptions of
the relevant singularities that are necessary to describe points of
enhanced gauge symmetry. 
It will turn out that two types of branes are sufficient to
describe the general situation. We shall also see that there exist
interesting partial resolutions of the $D_4, E_6, E_7$ and $E_8$  
singularities at constant $\tau$; these will prove very useful
for our explicit analysis of transitions.   

The most efficient way to describe $\tau$ as a function of $z$, is to
consider the torus bundle over $S^2$ that defines the K3 surface and
that is determined by the function
\be
\label{torus}
y^2 = x^3 + f(z)\,  x  + g(z) \,.  
\ee
Here $f$ and $g$ are polynomials in $z$ of degree eight and twelve,
respectively, and for each fixed $z$, (\ref{torus}) defines a torus
whose modular parameter $\tau$ can be defined implicitly as 
\be
j(\tau(z)) =  4 {(24f(z))^3 \over (4f(z)^3+27g(z)^2)} \,,
\ee
where $j(\tau)$ is the standard $j$ function.

It is not difficult to see that $\tau$ is constant if 
$f^3 \sim g^2$ \cite{senorientifold}, or if $g=0$ (branch I) and $f=0$
(branch II) \cite{dasgupta}. On the former branch, $g(z)\equiv 0$, and
$f(z)$ is an arbitrary polynomial of degree eight whose zeros coincide
with the zeros of the discriminant $\Delta(z)=4f(z)^3+27g(z)^2$
\be
f(z)  =  \prod_{i=1}^{8}(z-z_i) \,, \quad
\Delta(z)  =  4\prod_{i=1}^8 (z-z_i)^3 \,.
\ee
On this branch $\tau\equiv i$, and each of the eight zeros corresponds
to a singularity of type $A_1$ with fiber type III \cite{bikmsv}; this
gives rise to the enhanced symmetry $SU(2)^8$ at a generic
point.\footnote{Here and in the following, we do not write the
relevant $U(1)$ factors.} In the type IIB picture there are $24$
7-branes altogether, and thus we have to have three branes at each of
the eight singularities. Since the gauge algebra is $A_1$ (rather than
$A_2$) we have to conclude that two of the three branes are of the
same type, and that the third is relatively nonlocal to the former
two; one solution is given as   
\be
{\bf AAC}:  \qquad K_A K_A K_C = S \,,
\ee
which is indeed compatible with $\tau=i$, since this is a fixed point 
of $S$. The resulting metric is flat except at the singularities,
which are in this case conical, each with a defect angle of $\pi /2$.

On branch II, $f(z)\equiv 0$ and $g(z)$ is an arbitrary polynomial of
degree twelve whose zeros coincide again with those of the discriminant
\be
g(z)  =  \prod_{i=1}^{12}(z-z_i)\,, \qquad
\Delta(z)  =  27\prod_{i=1}^{12} (z-z_i)^2 \,. 
\ee
On this branch $\tau\equiv \exp (i\pi /3 )$, and each of the twelve
zeros corresponds to a fiber type II \cite{bikmsv}, which does
not give rise to an enhanced gauge symmetry. In the type IIB picture,
each of the twelve singularities are therefore realized by two
relatively non-local 7-branes; a solution can be given as 
\be
{\bf AC}: \qquad 
K_AK_C = TS \,,
\ee
which is consistent with constant $\tau\equiv \exp (i\pi /3 )$ since
this is a fixed point for the transformation $TS$.
In this case the singularities define conical singularities with a  
defect angle of $\pi/ 3$. 
\medskip

We can also understand from this point of view how the various other
enhanced symmetries arise as some of the singularities coincide. Let
us first consider branch I: as two of the {\bf AAC} collapse, the
singularity gets enhanced from 
$A_1\times A_1$ to  $D_4$. Indeed, in terms of the branes,
\be
\label{dfourm}
{\bf AACAAC} \longrightarrow {\bf AACBAA} \longrightarrow {\bf
CBAAAA}\,,  
\ee
which is recognized as the standard brane description of the $D_4$ 
singularity. In the first and second steps we used \myref{ftrans} and 
\myref{strans}, respectively.

When three bunches collide, the singularity becomes of $E_7$ type
\cite{bikmsv}. Using \myref{dfourm} this can be represented as 
\be
\label{dninem} 
{\bf AAC~AACAAC} 
\longrightarrow {\bf AAC~ CB AAAA} \longrightarrow 
{\bf C''C''B'' ~AAAAAA}\,,
\ee
where $K_{C''} = K_A^2 K_C K_A^{-2}$ and $K_{B''} = K_A^2 K_B K_A^{-2}$.
We thus recover (up to an irrelevant global $\mbox{SL}(2,\bbbz)$
transformation taking {\bf C}$''\to${\bf C}, {\bf B}$''\to$ {\bf B}
and {\bf A}$\to$ {\bf A}) the picture that was used in
\cite{johansen,GZ}.   
The coincidence of four bunches leads to ord$(f)=4$, which according
to \cite{bikmsv} destroys the triviality of the canonical bundle. 
It is not difficult to list all the possible enhanced gauge symmetries
that can be found in this way on branch I
\bean
\begin{array}{lll}
SU(2)^8   & E_7\times SU(2)^5 & E_7\times E_7\times SU(2)^2\\
SO(8)\times SU(2)^6      & E_7\times SO(8)\times SU(2)^3 & 
                             E_7\times E_7\times SO(8)\\
SO(8)^2\times SU(2)^4    & E_7\times SO(8)^2\times SU(2) &\\
SO(8)^3\times SU(2)^2    & & \\
SO(8)^4                  && \\
\end{array}
\eean
\smallskip

On branch II, the coincidence of two singularities gives rise to a
fiber type IV, giving an $A_2$ singularity \cite{bikmsv}. From the
point of view of the explicit branes this corresponds to 
\be
\label{su3}
{\bf ACAC} \longrightarrow {\bf AACA} \longrightarrow 
{\bf BAAA} \,.
\ee
The first step corresponds to moving the cuts of the left {\bf ACA}
block to the right of the last {\bf C} brane in three steps, one at a
time. This turns the {\bf C} brane into an {\bf A} brane
by virtue of $K_AK_CK_AK_CK_A^{-1}K_C^{-1}K_A^{-1} = K_A$. In the second 
step we simply used \myref{ftrans}.

Three coinciding singularities again give
rise to a $D_4$ singularity 
whose usual representation can now be recovered as 
\be
\label{so8}
{\bf AC~ACAC} \longrightarrow {\bf AC~BAAA} 
\longrightarrow {\bf CBAAAA}\,,
\ee
where we have used \myref{su3} in the first step, and \myref{strans}
in the second step.    
Four coinciding singularities result in an $E_6$
singularity. 
Using (\ref{so8})
this can be understood as 
\bean
{\bf AC~ACACAC} \longrightarrow {\bf AC~CBAAAA} 
\longrightarrow {\bf C' C' B' AAAAA}\,,
\eean
with $K_{C'}=K_AK_CK_A^{-1}$,   
and  $K_{B'}=K_AK_BK_A^{-1}$.   
This can be recognized as the description used in
\cite{johansen,GZ}. Finally, when five singularities coincide,
we get an $E_8$ singularity.   
Using (\ref{su3}) and (\ref{so8}) this corresponds to
\bea
{\bf ACAC~ACACAC} \longrightarrow 
& {\bf AACA~ CBAAAA}  \longrightarrow 
{\bf AA~CCB~AAAAA} \longrightarrow \nn \\
& {\bf C'' C'' B'' ~AA~AAAAA} \,.
\eea
This reproduces the description
of \cite{johansen,GZ}. The allowed symmetry enhancements on branch II
are therefore 
\bean
\begin{array}{lllll}
0 & SO(8)&SO(8)^2&SO(8)^3&SO(8)^4\\
SU(3)&SO(8)\times SU(3)&SO(8)^2\times SU(3)&SO(8)^3\times SU(3)&\\
SU(3)^2&SO(8)\times SU(3)^2&SO(8)^2\times SU(3)^2&&\\
SU(3)^3&SO(8)\times SU(3)^3&SO(8)^2\times SU(3)^3&&\\
SU(3)^4&SO(8)\times SU(3)^4&&&\\
SU(3)^5&&&&\\
SU(3)^6&&&&
\end{array} 
\eean
\bean
\begin{array}{llll}
E_6& E_6\times SO(8) & E_6\times E_6& E_6\times E_6\times E_6\\
E_6\times SU(3)& E_6\times SO(8)\times SU(3)&E_6\times E_6\times SU(3)&\\
E_6\times SU(3)^2& E_6\times SO(8)\times SU(3)^2 &
E_6\times E_6\times SU(3)^2&\\
E_6\times SU(3)^3&E_6\times SO(8)^2   &E_6\times E_6\times SO(8) \\
E_6\times SU(3)^4&E_6\times SO(8)^2 \times SU(3) & &\\
\end{array} 
\eean
\bean
\begin{array}{llll}
E_8&E_8\times SO(8)&E_8\times E_6 &E_8\times E_8 \\
E_8\times SU(3) &E_8\times SO(8)\times SU(3)&
E_8\times E_6 \times SU(3) & E_8\times E_8 \times SU(3) \\
E_8\times SU(3)^2 &E_8\times SO(8)\times SU(3)^2&
E_8\times E_6 \times SO(8) & \\
E_8\times SU(3)^3 &E_8\times SO(8)\times SO(8)&&
\end{array}
\eean

\section{Transitions in Constant $\tau$ Backgrounds}
\label{sec:constant}
\setcounter{equation}{0}

{}From the point of view of F-theory, the relevant BPS states
correspond to supersymmetric cycles in an elliptically fibered K3
whose self-intersection numbers are $-2$. In type IIB theory these
states are described by smooth geodesics or possibly string
junctions on the base space of the elliptic fibration (a two-sphere)
whose metric is given by (\ref{effmet}). $h_{p,q}(z)$ is an analytic 
function of $z$ save for possible poles at the locations of the
7-branes, and the metric is therefore flat except  at the
locations of the poles.

The relevant smooth geodesics in type IIB begin and end at the
position of a 7-brane, and, as they are smooth, must avoid the
singularities of the other 7-branes. Each such geodesic defines
therefore an element of a homotopy class. The property to be geodesic
means that its length is minimal in the homotopy class of curves it
defines. The computation of the length may require the use of several
metrics, as the $({p\atop q})$ labels of the string can change
whenever it crosses a cut. Given a fixed set of branch cuts, we can
distinguish between homotopy classes associated to {\it direct
strings}, containing representatives that do not cross branch cuts,
and homotopy classes associated to {\it indirect strings}, where all
representatives must necessarily cross branch cuts.

Given a homotopy class with two fixed endpoints, we can analyze
whether a smooth geodesic exists in this class. If this is not the
case, we can ask whether there exists a geodesic string junction or 
some other geodesic that corresponds to the same cycle in K3. (In
particular, the configurations that can be obtained by brane crossings
always define the same cycle in K3.) Here a geodesic three-string
junction is a trivalent graph, where the three prongs are smooth and
end on 7-branes. The total length of the geodesic junction is
defined as the sum of the lengths of the prongs. It is not 
difficult to show that in order for the string junction to be geodesic
({\it i.e.} of minimal length), the familiar force balance condition
\cite{schwarzreview,sennetwork} must be satisfied at the junction
vertex. 

We shall find (at least in the examples we consider) that for the
cycles with self-intersection number $-2$, {\it precisely one} of the
possible geodesics objects is smooth and thus represents the BPS
state in this configuration. This confirms the proposal of \cite{GZ}
that the conventional geodesics are not the relevant BPS states in the
whole of the moduli space; it also reaffirms indirectly, via a string
of dualities, the Hanany-Witten effect \cite{hananywitten} (see also
\cite{OV}), and the string creation effect in the D0-D8 brane system
\cite{08}.  
\smallskip

In this section we use the resolutions of the various gauge
enhancement points (that have been discussed in the previous section)
for which $\tau$ is constant. This will simplify the analysis
considerably, as there exists then a coordinate in which {\it all}  
$(p,q)$ geodesic strings are straight lines. We shall first consider
the resolution of $so(8)$ into two bunches of 7-branes, and we shall
see that an indirect {\bf A--A} geodesic is in fact realized as a
degenerate three string junction. Much of the technology shall be
developed through this example. We shall then turn to the case of a
three singularity resolution of $so(8)$ where the transitions between
a direct {\bf A-A} geodesic, the corresponding {\bf A-A-C} geodesic
junction, and the associated indirect {\bf A-C} strings can be seen
very explicitly. We also discuss briefly other examples. The main
lesson of this section is that in the appropriate coordinate the total
length of a string  junction is given by the length of a mathematical
straight line joining the positions of the initial and final
7-branes.

\subsection{Indirect strings in a two-singularity 
resolution of ${\bf so(8)}$}
\label{sec:twosing}

Let us first consider the $so(8)$ singularity which we resolve at
constant $\tau=i$ as ({\bf AAC}$\quad$ {\bf BAA}). For definiteness,
let us place the {\bf AAC} group at $z_0$ and the  {\bf BAA} at $z_1$,
where both $z_0=x_0$ and $z_1=x_1$ are real and  $x_0< x_1$
(Fig.~\ref{fig4}(a)). 
\onefigure{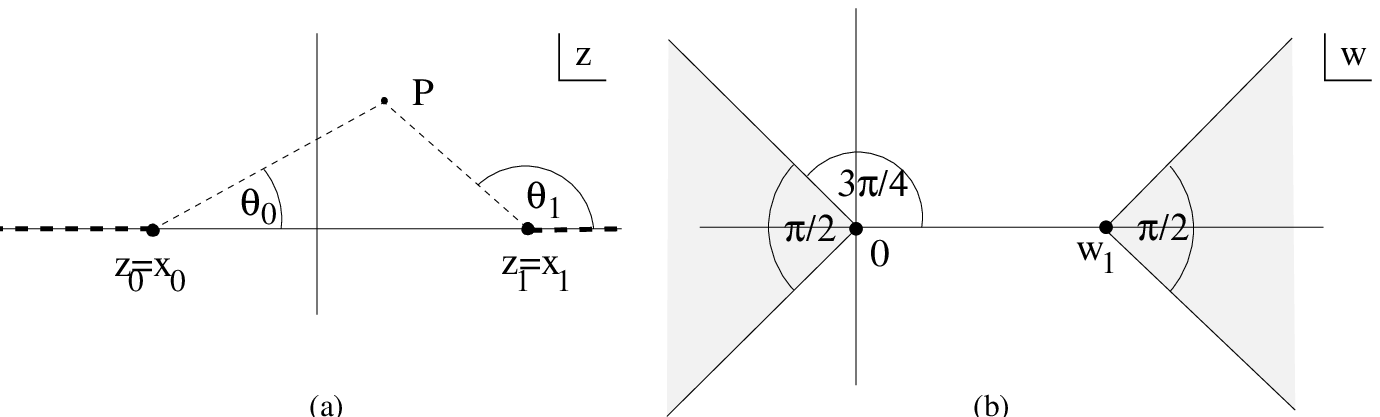}{Two-singularity resolution of so(8). (a) The groups
     of three branes are located at the real points $z_0$ and $z_1$.
     (b) The singularities are mapped to $0$ and $w_1$ while the whole 
     $z$-plane is mapped to the non-shaded region of the $w$-plane.
     \label{fig4}} 

The metric is defined as
\bean
\label{wpictwo}
h_{p,q}(z)&=& C (p - iq) (z-z_0)^{-\frac{1}{4}}(z-z_1)^{-\frac{1}{4}}\,,
\eean
where $C$ is some constant that does not depend on $(p,q)$. All
strings feel the same identical metric up to an overall $(p,q)$
dependent constant, and thus all geodesics are necessarily along the
same trajectories. The metric is flat and has conical type
singularities of deficit angle $\frac{\pi}{2}$ at $z_0$ and
$z_1$. 

Let us take $(p,q) = (1,0)$, $C= \exp (i\pi/4)$, and let us draw a
branch cut running horizontally to the left from $z_0$ with angles
defined by $-\pi \leq \theta_0 \leq \pi$  on the sheet, and a second
cut running horizontally to the right from $z_1$ with angles defined
by $0 \leq \theta_1 \leq 2\pi$. Let us introduce a new coordinate $w$
via 
\be
\label{wnew}
w(z) = \int_{z_0}^z {e^{i\pi /4}\,  dz'\over 
[(z'-x_0)(z'-x_1)]^{\frac{1}{4}} } \,.
\ee
By construction it is clear that in the Euclidean $w$-plane, geodesics
are represented by straight lines. The singularity at $z_0$ maps to a
singularity at $w=0$, and the two sides of the branch
cut emanating from $z_0$ are mapped to rays with arguments 
$\pm 3 \pi/4$ in the $w$-plane (see Fig.~\ref{fig4}(b)). The other
singularity at $z_1$ is mapped to $w_1$ which is real because
\bean
w_1= \int_{x_0}^{x_1} {dx\over 
[(x-x_0)(x_1-x)\,]^{\frac{1}{4}}} \,.
\eean
The two sides of the cut originating at $z_1$ are mapped to the rays
departing from $w_1$ with arguments $\pi/4$ and $2\pi - \pi/4$. It
is then clear that the image of the whole cut $z$-plane is the part of the
$w$-plane, where appropriate wedges with $|\theta_{0}| > 3\pi/4$ at
$w=0$, and $|\theta_{w_1}|\leq \pi/4$ at $w_1$ have been removed; this
is depicted in Fig.~5(b). There is an implicit identification of
the rays representing the boundaries of the shaded regions, but this
should not be taken to mean that the metric is simply an ordinary
conical metric; in fact, the type of string changes as it crosses the
seam described by the identification.

The structure of the $w$-plane could have been guessed directly from
inspection of \myref{wnew}, and the only nontrivial information
that the map carries is a determination of $w_1$ in terms of $x_1$ and
$x_0$. As we consider $w_1$ to be an adjustable parameter, its
relation to the $z$-plane parameters is not important for us and we
can work directly in the $w$ plane.  This will also be the case for  
metrics with three singular points.
\smallskip

We can consider resolving the two bunches of branes further into
individual branes (see Fig.~\ref{fig5}(b)), so that we can distinguish
more 
clearly the various types of geodesics, but this is not compatible
with the assumption of constant $\tau=i$. We shall therefore use this
further resolution only as a formal tool, with the implicit
understanding that we have to take the limit in which the individual
branes collapse to the above bunches. We want to show that (in this
limit) the indirect {\bf A-A} geodesic is not smooth and that the
associated three-string geodesic junction {\bf A-A-C} has total length
which is smaller than that of the indirect  {\bf A-A} geodesic by a
finite amount proportional to the distance between the two
bunches. This should be sufficient to guarantee that the three-string 
junction remains the actual BPS state for a small resolution of the
bunches into individual branes.
\onefigure{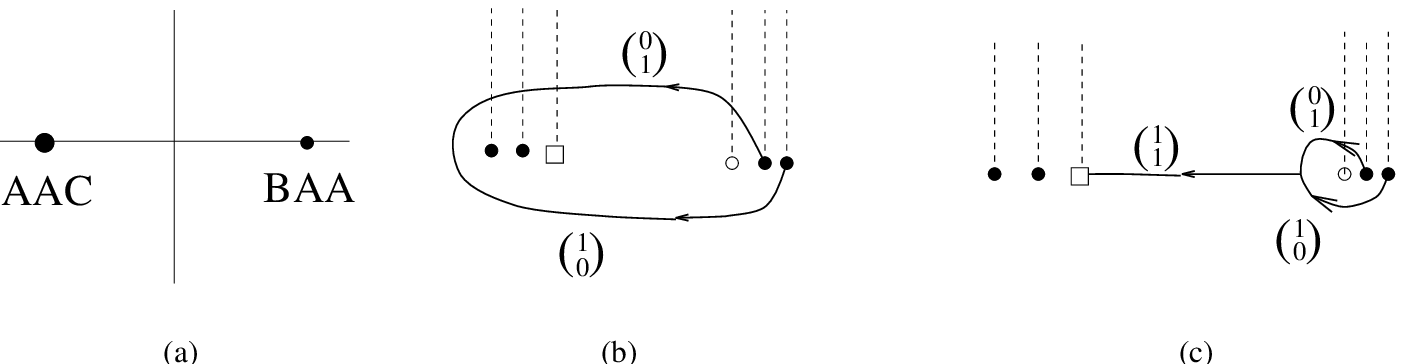}{(a) The two-singularity resolution of so(8). (b)
     An indirect {\bf A-A} string. (c) The corresponding three-pronged
     junction. When the branes eventually collapse
     into two punctures the $\vect{0}{1}$ and the $\vect{1}{0}$ prongs
     degenerate. \label{fig5}}

The indirect {\bf A-A} string departs from each of two {\bf A}-branes
in the {\bf AAB} bunch as a $({1\atop 0})$ string, and it is a 
$({1\atop 0})$ and $({0\atop 1})$ string, respectively in between the
bunches (see Fig.~\ref{fig5}(b)). As the bunches collapse, the top and
bottom 
geodesic strings are straight lines from one bunch to the other, and
the total length is thus the sum of the two contributions
\bean
|{\bf \hbox{A\,-A}}| = l_{1,0} + l_{0,1} = 2 w_1 \,.
\eean
This path is not a smooth geodesic: as the path goes through the
singularity at $z_0$ it reverses direction.

This geodesic should be compared with the junction that corresponds to
the same cycle in K3. This junction arises as the {\bf A-A} string
cuts through the {\bf C} brane in the left singularity producing a
{\bf C} prong, {\it i.e.} a $({1\atop 1})$ string (see 
Fig.~\ref{fig5}(c)). 
As the resolved branes recollapse into the two separate bunches we get
$({1\atop 0})$ and  $({0\atop 1})$ prongs of negligible length, but a
long $({1\atop 1})$  prong connecting the two bunches, and thus
\bean
|{\bf \hbox{A\,-A\,-C}}| = \sqrt{2}\, w_1 \,,
\eean
where the $\sqrt {2}$ arises as the absolute value of the
$(1-i)$ factor in the metric. It thus follows that
\bean
{|{\bf \hbox{A\,-A\,-C}}|\over |{\bf \hbox{A\,-A}}|} = 
{1\over \sqrt{2}} \,,
\eean
and we have verified that the geodesic junction has lower length than
the non-smooth open string geodesic. Because of the above arguments
this result should also hold when the two bunches are resolved
infinitesimally.

\subsection{Transitions for a three singularity
resolution of so(8)} 
\label{sec:threesing}

In the above example, the 3-pronged string was somewhat degenerate in
that two of the prongs had vanishing length in the limit where
the branes collapsed to bunches. We shall now consider examples where
this does not happen. To this end we have to consider resolutions (at
constant $\tau$) into at least three bunches of branes. The simplest
case already occurs (perhaps slightly surprisingly) for $so(8)$ which
can be resolved as 
\bean{ {{\bf  AC}\atop P_1} \hspace{1in} {{\bf AC}\atop Q}
\hspace{1in}  {{\bf AC}\atop P_2} 
}\eean
where $\tau = \exp (i\pi/3)$ (see Fig.~\ref{fig6}). We expect that as
long as the middle singularity is above the line between the two other
singularities, there exists a smooth geodesic representing a direct
{\bf A-A} string connecting the right and left bunches of branes. As
the middle singularity is moved downwards, we expect that the smooth
(direct) geodesic eventually develops a corner, and that there is a
transition to a geodesic three-pronged junction whose different prongs
end on the three singularities.
\onefigure{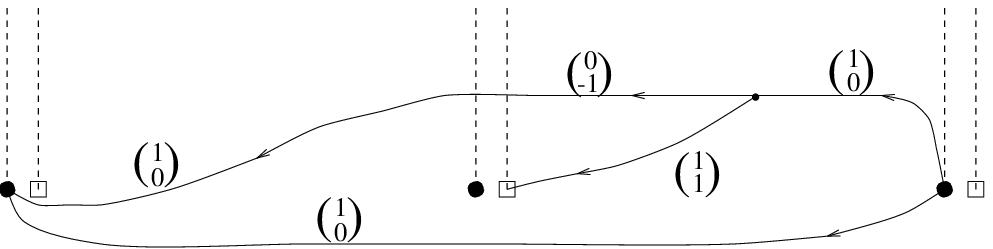}{The direct {\bf A-A} string and the 
     corresponding three-pronged {\bf A-A-C} junction in the
     three-singularity resolution of so(8). \label{fig6}}

It is again useful to resolve the three bunches further in order to 
identify the types of strings (Fig.~\ref{fig6}). The direct string is
a $(1,0)$ string that avoids any branch cut, and the three-pronged
string starts as a $(1,0)$ string on the right {\bf A}-brane, emits a
$(1,1)$-string (that ends on the middle {\bf C} brane), and thus
becomes a $(0,-1)$ string. It then crosses the {\bf AC}-cut thereby
becoming a $(1,0)$ string that can end on the {\bf A}-brane of the
left singularity (avoiding the leftmost {\bf C}-cut). The
configuration in the $z$-plane is depicted in 
Fig.~\ref{fig7}(a) where the
location of the three bunches are labeled as $P_1$, $Q$ and $P_2$,
the position of the junction is denoted by $S$, and $B$ and $A$ are
the points to the right and left of the branch cut emanating from $Q$.

As before we introduce the coordinate
\bea
\label{ztow}
w(z)=\int^z_{z_{P_2}} h_{1,0}(z')dz'\,, \quad 
h_{1,0} (z) = C(z-z_{P_1})^{-\frac{1}{6}}
(z-z_{P_2})^{-\frac{1}{6}}(z-z_{Q})^{-\frac{1}{6}}\,,
\eea
that is appropriate for $(1,0)$ strings. The two sides of the cut
emanating from $Q$ become rays that enclose an angle of $60^\circ$,
and the image of the cut $z$-plane is the $w$-plane where the shaded
region has been removed (see Fig.~\ref{fig7}(b)).\footnote{The precise
form of the shaded region depends on the choice of the branch cut that
emanates from $Q$ in the $z$-plane.} As $\tau=e^{i\frac{\pi}{3}}$ we
have that $T_{1,0}=T_{1,1}=T_{0,-1}$, and thus the appropriate length
for the corresponding strings coincides with their actual length in
the $w$-plane (up to an overall immaterial factor).
\onefigure{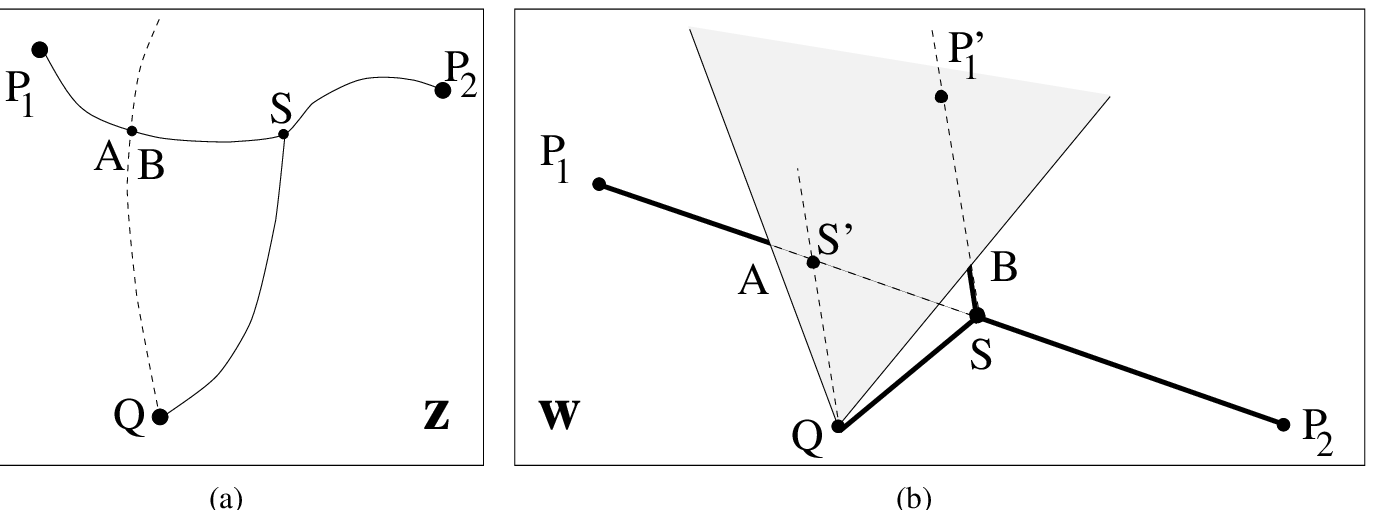}{(a) Three-pronged junction in the $z$-plane. (b)
     Three-pronged junction in the $w$-plane. Since
     $\tau=e^{i\frac{\pi}{3}}$, the $w$-plane represents faithfully
     the effective length of the prongs. The total mass of the
     junction is given by the length of the straight segment $P_1P_2$.
     \label{fig7}} 

As long as the $P_1 P_2$ line segment in the $w$-plane avoids the
shaded region, the corresponding geodesic is smooth. If this is not
the case (as shown in Fig.~\ref{fig7}(b)), the shortest path that
avoids the shaded region (and thus corresponds to an element in the
same homotopy class as before) goes through the singularity at $Q$,
and thus does not represent a smooth geodesic. On the other hand, the 
three-string junction that is depicted in Fig.~\ref{fig7}(b) is a
smooth geodesic and its overall length is strictly smaller than the
length of the above path. Let us justify this claim in some detail. 

First, in order to see that the diagram represents a geodesic junction
we observe that (i) the angles at the junction $S$ are all
$120^\circ$, and (ii) all prongs are straight lines. (This is obvious
for the $SQ$ and $SP_2$ prongs; as the defect angle at $Q$ is
$60^\circ$, the $SP_1$ is prong is a straight line provided that the
$AP_1$ line when rotated by $60^\circ$ to $BP_1'$ forms a straight
line with the first portion $SB$.)

In fact, given $Q$, $P_2$ and $P_1'$, there exists at most one point
$S$ for which these conditions are satisfied, and this point can
be found as follows. The points $X_1$ for which 
$\angle QX_1P_1' = 120^\circ$ form an arc (which goes through $Q$ and
$P_1'$), and so do the points $X_2$ for which 
$\angle QX_2P_2 = 120^\circ$. $S$ is therefore the (unique)
intersection of the two arcs. (The intersection point may not
exist, in which case the junction does not exist, see below.) The line
$P_1'S$ determines then the point $B$, and thus also $A$.

Next, we want to determine the overall length of this (unique)
junction. As the triangle $QP_1 P_1'$ is equilateral, $P_1$ is on the 
arc over $QP_1'$ with angle $60^\circ$, and $S$ is on the
arc over  $QP_1'$ with angle $120^\circ$, and thus the two arcs 
form together an actual circle. This implies that both $Q$ and $S$ lie
on the same arc over $P_1 P_1'$ with angle $60^\circ$, 
and therefore that the angles 
$\angle P_1SP_1'= \angle P_1QP_1' = 60^\circ$. It then follows 
that the points $P_1$, $S$ and $P_2$ define a line, and thus that
the $P_1 A$ and $P_2 S$ prongs lie on the $P_1P_2$ line segment. 

Let us denote by $S'$ the image of $S$ after an counterclockwise
rotation by $60^\circ$ around $Q$. As this rotation maps B to A, it
then follows that $AS'Q$ and $BSQ$ are identical triangles. Finally,
since $QSS'$ is equilateral, the distance $SQ$ equals $SS'$, and we
therefore find that the {\em length of the $P_1P_2$ line segment is
equal to the true length of the junction}. In particular, this implies
that the length of the junction (if it exists) is strictly lower than
that of the broken direct geodesic. Conversely, if the direct geodesic
is smooth, $Q$ must lie above the  $P_1P_2$ segment, and the above
construction leads to two arcs that do not intersect, and 
the geodesic junction does not exist.
\medskip

It is actually possible to establish the existence of the junction
directly by choosing the shaded region to lie symmetrically with
respect to the line passing through $Q$ and perpendicular to
$P_1P_2$. (This choice corresponds to a certain choice for
the location of the branch cut that emanates from $Q$.)
In this case, the length of the $(0,-1)$ string is zero, and
there are no broken segments in the picture (Fig.~\ref{fig8}).
\onefigure{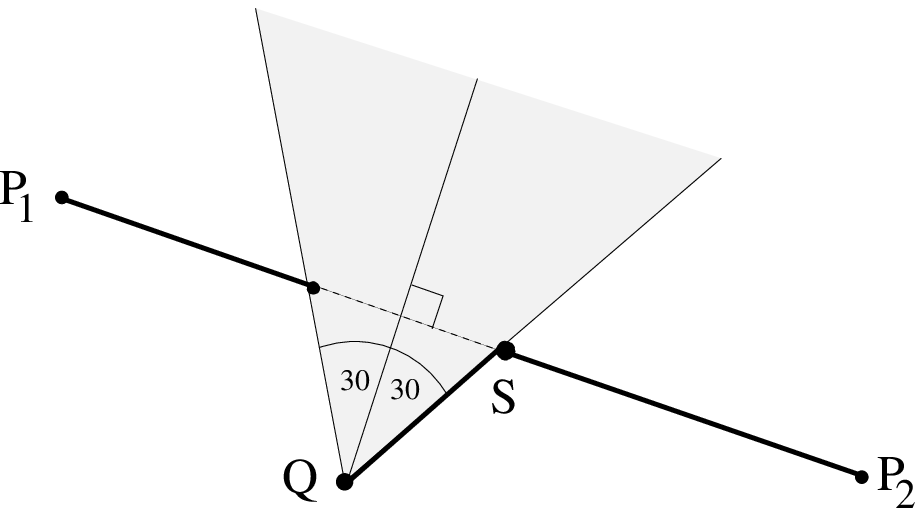}{In the $w$-plane the junction simplifies if the
     cut is chosen to lie along the $QS$ prong. \label{fig8}}
It is then very easy to visualize the open string/junction
transition. Let us fix the points $P_1$ and $P_2$ on the $w$-plane and
suppose that $Q$ is somewhere above the $P_1P_2$ segment as in
Fig.~\ref{fig9}(a). As $Q$ is moved downwards and crosses the $P_1P_2$
line, the would-be geodesic string is transformed into a geodesic
junction (Fig.~\ref{fig9}(b) and (c)). It should be stressed that the
mass of the representative of this BPS state is unchanged in the
process, as it is given by $\overline{P_1P_2}$.
\onefigure{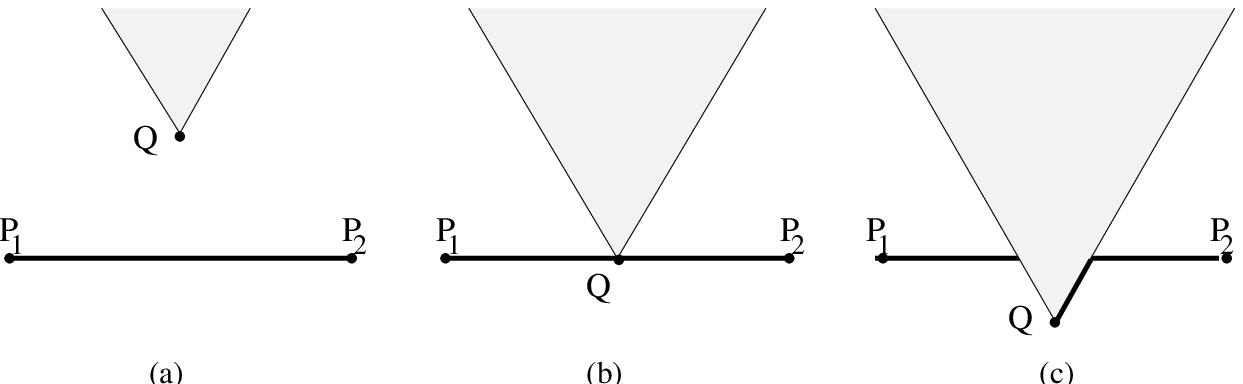}{For fixed $P_1$ and $P_2$, a prong is created as
     the third puncture $Q$ crosses the $P_1P_2$ line segment.  
     \label{fig9}}
\bigskip

This argument covers most of the possible configurations, but there
exists yet another region where $Q$ is below the $P_1 P_2$ segment (so
that the direct geodesic has a corner), but yet the angle 
$\angle \, P_1'P_2Q > 120^\circ$, and thus we cannot find a
satisfactory junction $S$ as the relevant arcs do not intersect 
(Fig.~\ref{fig7}(b)). This configuration can be reached from the
three-string configuration discussed above as $P_2$ approaches $S$,
and the length of one of the prongs of the junction vanishes. We want
to demonstrate next that as $P_2$ crosses $S$, the junctions turns
into an indirect {\bf C-A} string that runs from $Q$ to $P_1$
(Fig.~\ref{fig14}). 
% \onefigure{fig14.eps}{The indirect {\bf A-C} string. 
% The junction of Figure.~7 can be obtained by moving the string across
% the rightmost singularity. \label{fig14}}
\onefigure{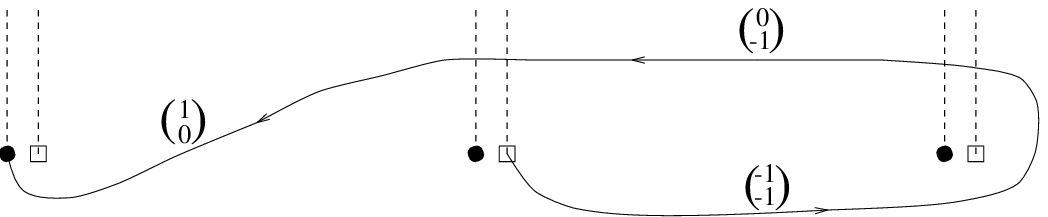}{The indirect {\bf A-C} string. 
     The junction of Fig.~\protect\ref{fig6}              % Figure.~7 
     can be obtained by moving the string across
     the rightmost singularity. \label{fig14}}

In order to discuss this configuration, it is necessary to draw also
the branch cut that emanates from $P_2$, which in the $w$-plane again
corresponds to a wedge of angle $60^\circ$ that is drawn as a shaded
region in Fig.~\ref{fig15}. Let $P_1'$ be the image of $P_1$ after a
rotation about $Q$ by $60^\circ$, and let $P_1''$ and $Q'$ be the
images of $P_1'$ and $Q$ after a rotation about $P_2$ by  
$60^{\circ}$ and $-60^{\circ}$, respectively. The relevant (indirect)
geodesic is depicted in heavy lines, and it is manifest that it does
not have any corners. By rotating the line segments about $P_2$ and
$Q$, it is easy to see that the total length of the string 
equals $\overline{ P_1'Q'}$. Finally, since the two triangles
$P_1QP_2$ and $P_1'QQ'$ are identical (as the triangle $P_2 Q Q'$ is 
equilateral), it follows that the overall length is again equal to 
$\overline{P_1P_2}$. 

It should be noted that $\angle \, P_1'P_2Q > 120^\circ$ implies that
$Q'P_1'$ crosses the excised region emanating from $P_2$,
which is necessary for the charge conservation of the indirect
string. (The indirect string has to cross the cuts from both $Q$ and 
$P_2$, as is immediate from Fig.~\ref{fig14}.) Completely analogous
arguments hold also for the indirect {\bf A-C} string running from $Q$
to $P_2$.

\onefigure{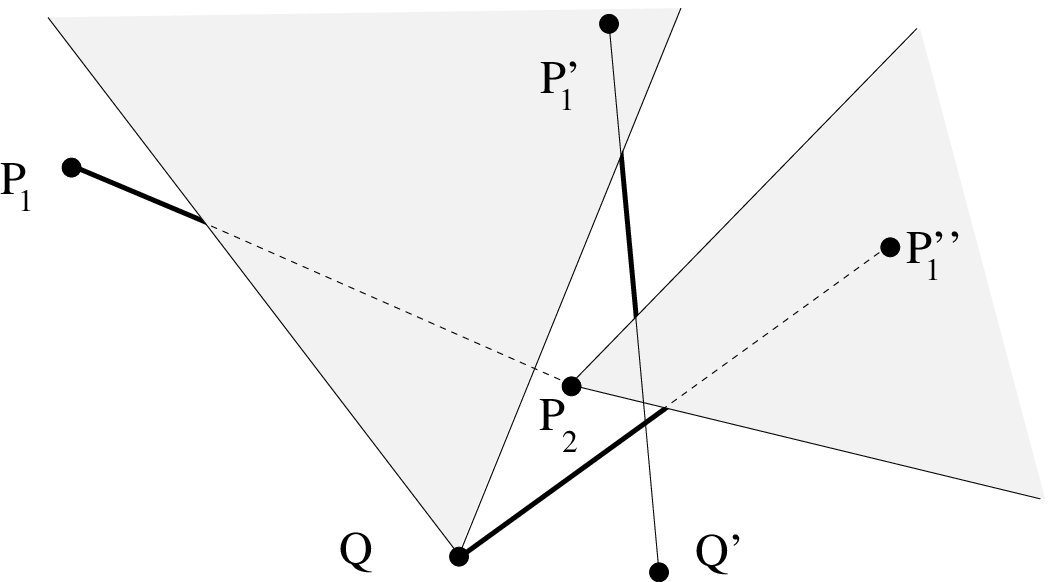}{The indirect {\bf A-C} string
     is represented by three broken line segments in the
     $w$-plane. Charge conservation requires that it crosses
     both the $P_2$ and the $Q$-cut. \label{fig15}}
\medskip
\onefigure{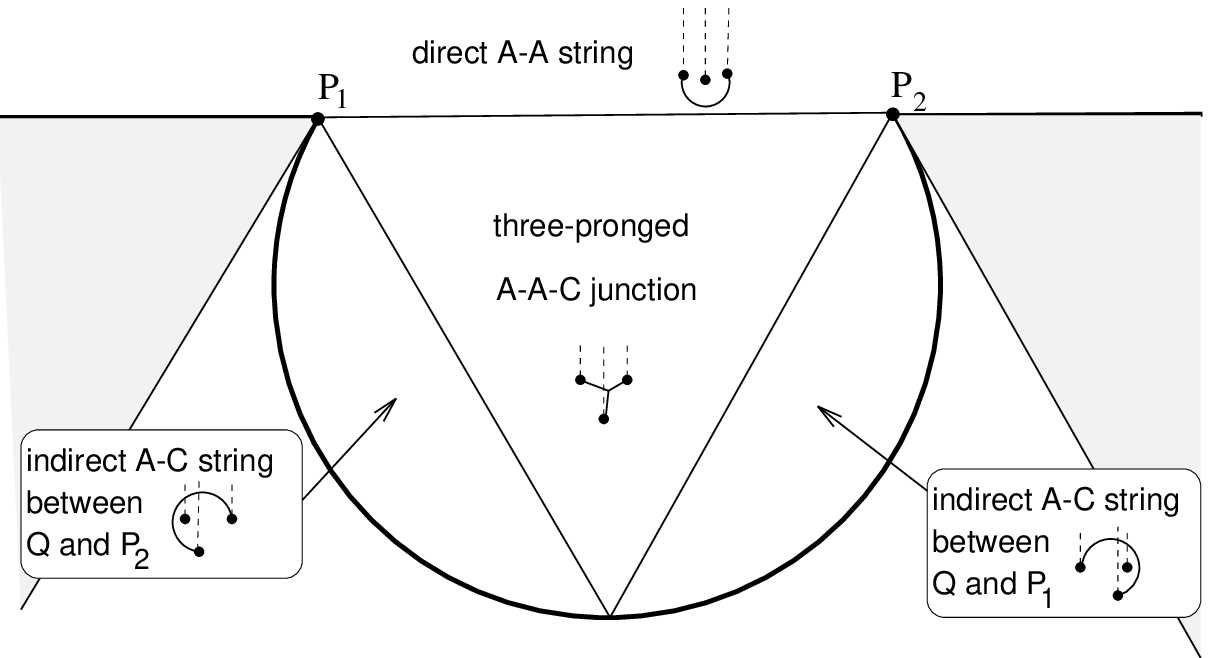}{For fixed $P_1$ and $P_2$ we have depicted the
     different regions for $Q$ in the $w$-plane where different
     configurations are BPS. The shaded regions are the excised wedges
     associated to $P_1$ and $P_2$, and the arc through $P_1$ and
     $P_2$ is the boundary of the region where both $P_1$ and $P_2$
     lie outside the wedge emanating from $Q$. \label{fig16}}
We have thus found that depending on the configuration of the
7-branes, the actual BPS state is realized either as a direct 
{\bf A-A} string, as one of two indirect {\bf C-A} strings, or as a
{\bf A-A-C} junction. Only one realization is smooth at any point of
the moduli space (see Fig.~\ref{fig16}), and its mass is always given
in terms of the length of the straight line joining the two 
{\bf A}-branes in the $w$-plane. As we shall demonstrate below (see
section~\ref{sec:nonconstant}), similar results also hold for the case
where $\tau$ is not constant.

\subsection{Other three-singularity resolutions}
\label{sec:moreconical}

We can also consider other resolution of singularities into three
bunches of branes. For example, we can resolve the $E_6$ singularity
at $\tau=\exp (i\pi/{3})$ as 
\bean{ {{\bf  AC}\atop P_1} \hspace{1in} {{\bf BAAA}\atop Q}
\hspace{1in}  {{\bf AC}\atop P_2} 
}\eean
The direct {\bf A-A} string and its junction counterpart are shown
in Fig.~\ref{fig11}(a).
The defect angle associated with $Q$ is $120^\circ$ in this case,  
and the junction is shown in Fig.~\ref{fig11}(b). The angles at $S$ are 
$150^\circ, 150^\circ$  and $60^\circ$, and the effective length of
the $QS$ $(-1,1)$-prong is indeed equal to 
the length of the portion of the line segment $P_1P_2$ inside the
excised region, by virtue of $T_{-1,1} = \sqrt{3}\, T_{1,0}$. Again
we see that the effective length of the junction equals the length of
the straight mathematical line joining $P_1$ and $P_2$.
\onefigure{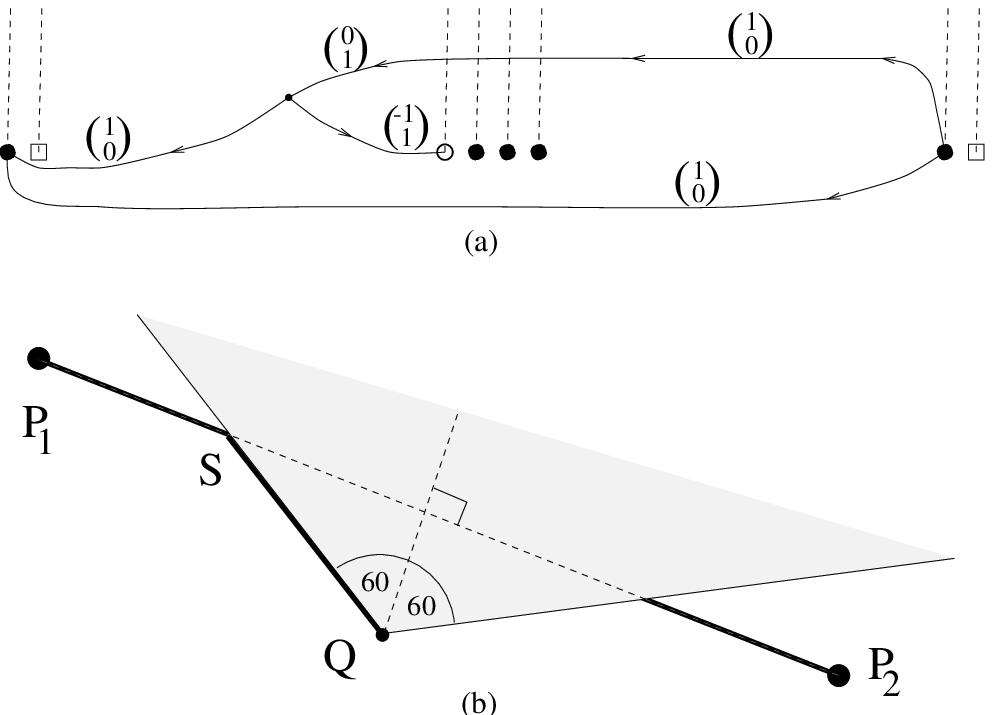}{(a) Constant $\tau$ three-singularity resolution
     of $E_6$. (b) The conical singularity at $Q$ ({\bf
     BAAA}) has defect angle $120^{\circ}$. The physical length of the
     $SQ$ prong of the geodesic junction is
     $\protect\sqrt{3}(\overline{SQ})$ which is equal to the length of
     the portion of $P_1P_2$ inside the shaded area.  \label{fig11}}  

There is also a three singularity resolution of the ten 7-branes that
make up the $E_8$ singularity as {\bf AC}$\quad$
{\bf ACBAAA}$\quad${\bf AC}, where $\tau=\exp (i\pi/3)$. The middle
singularity has a defect angle of $180^\circ$, and thus the direct
strings always remain geodesics.
\smallskip

On the branch $\tau=i$, there exists a resolution of the $E_7$
singularity as (see Fig.~\ref{fig13}(a))
\bean{  
{{\bf  AAC}\atop P_1} \hspace{1in} 
{{\bf AAC}\atop Q}  \hspace{1in} {{\bf AAC} \atop P_2}
}\eean
\onefigure{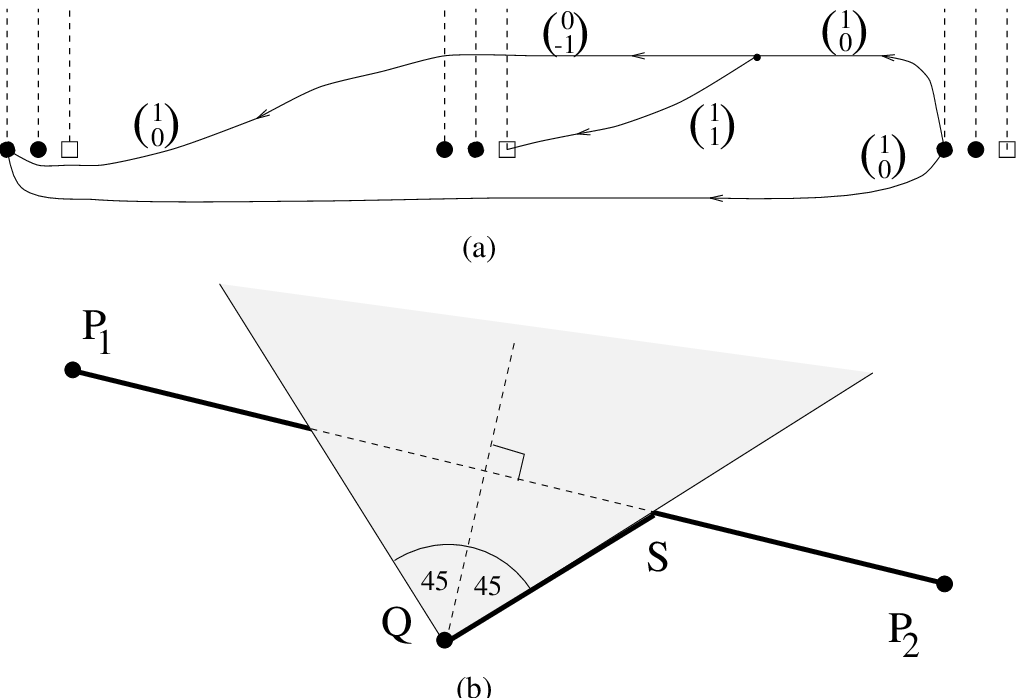}{(a) Constant $\tau$ three-singularity resolution
     of $E_7$. (b) The conical singularity at $Q$ ({\bf
     AAC}) has defect angle $90^{\circ}$. The physical length of the $SQ$
     prong of the geodesic junction is
     $\protect\sqrt{2}(\overline{SQ})$ which is equal to the length of
     the portion of $P_1P_2$ inside the shaded area. \label{fig13}}
The conical singularity at $Q$ now carries a defect angle of
$90^\circ$, and the effective length of the junction is given again by
the distance between $P_1$ and $P_2$, as 
$T_{1,1} = \sqrt{2}\,T_{1,0}$ for $\tau=i$ (Fig.~\ref{fig13}(b)).

\section{Geodesic junctions in general $\tau$-backgrounds}
\label{sec:nonconstant}
\setcounter{equation}{0}

In the previous section we gave a geometrical interpretation of the
length of geodesic prongs that are created as the corresponding open
string geodesic develops corners. In particular, we considered open
string geodesics between singularities $P_1$ and $P_2$ that fail to be
smooth when a third singularity $Q$ crosses the line $P_1 P_2$. Our
main observation was that in the $w$-plane, the length of the
$Q$-prong was represented as that part of the straight line segment
connecting $P_1$ and $P_2$ which falls into the wedge that has been
removed from the $w$-plane on account of the  cut originating
at $Q$ in the $z$-plane. We  
want to explain now that effectively the same mechanism also works in
the case where $\tau$ is not constant.

\onefigure{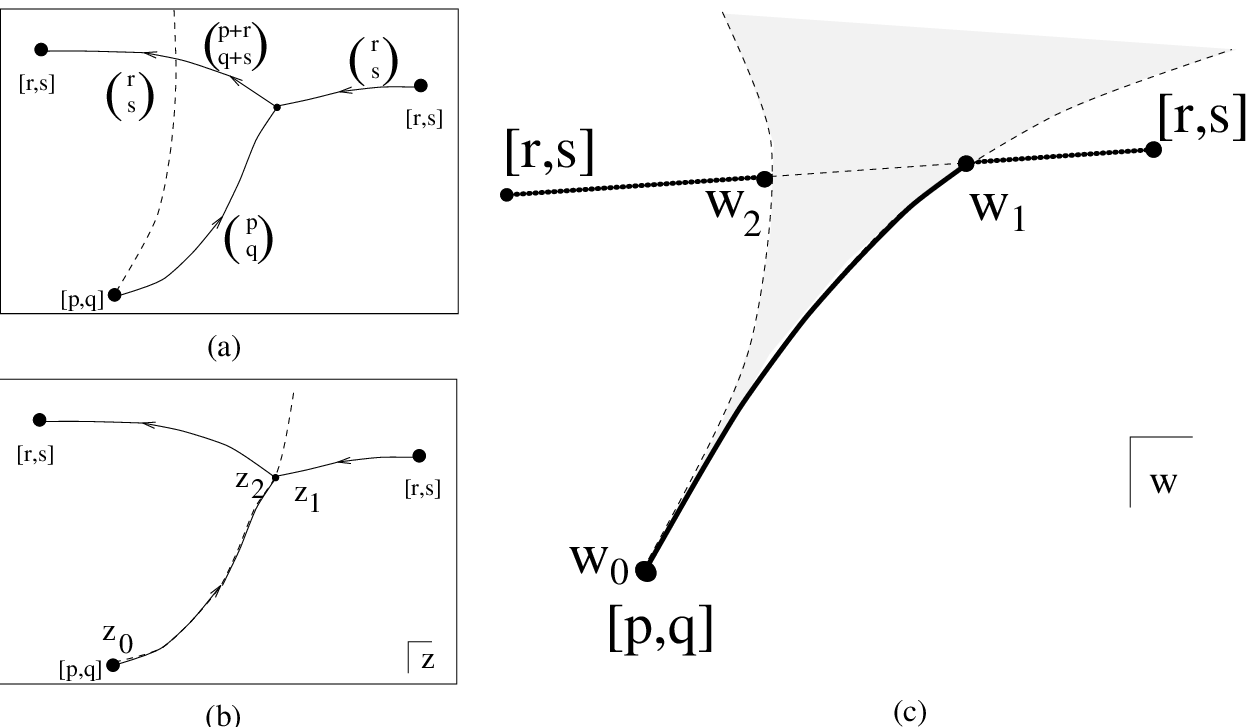}{(a) The three-string junction that is obtained 
     as the $({r\atop s})$-string between the two $[r,s]$-branes
     crosses the $[p,q]$ brane. (b) The geodesic three-pronged 
     junction with a convenient placement of the cut.  
     (c) In the $w$-plane the
     length of the geodesic $\vect{p}{q}$-prong is given by
     $\overline{w_1w_2}$ so that the mass of the state is still
     measured by the line segment between the two
     $[r,s]$-branes. \label{fig17}}
Let us consider an $\vect{r}{s}$-string in the vicinity of a
$[p,q]$-brane, to which it is compatible, say,  $ps-qr=1$\footnote{This
compatibility condition is equivalent to demanding that the associated 
process of brane crossing creates a single $({p\atop q})$ prong.} (see 
Fig.~\ref{fig17}(a)): 
a $\vect{p}{q}$-string and a $\vect{r}{s}$-string meet at the
junction, whose third prong has quantum numbers $\vect{r+p}{s+q}$. 
This prong crosses the cut of the $[p,q]$-brane counterclockwise and
is transformed back to an $\vect{r}{s}$-string. Charge conservation
requires 
\bea
\label{chcons}
M_{p,q} \pmatrix{r\cr s} = \pmatrix{ r+p\cr s+ q}\,,
\eea 
which holds if $ps-qr=1$.

Although all $(p,q)$ metrics are flat away from branes, when $\tau$ is
not constant there is no coordinate in which all $({p \atop q})$
strings are straight lines. Given that two of the prongs ultimately
are $({r\atop s})$ strings it is useful to introduce the
$w$-coordinate where such strings are straight lines
\bea
w(z)&=& \int^zh_{\vect{r}{s}}(z)\, dz\, .
\label{wrs}
\eea
In general, the geodesic of the $\vect{p}{q}$-prong is then a curved
line in this $w$-plane, whose explicit form may be difficult to
determine. However, as we shall show, its length can be easily
visualized in the $w$-plane.

The $\vect{r}{s}$ metric and the corresponding $w$ coordinate has a
cut emanating from the $[p,q]$-brane (whose position we denote by
$z_0$ in the $z$-plane and $w_0$ in the $w$-plane). 
The precise position of the cut is not physical, and we are free
to choose it when defining $w$. Suppose we are given already the
physical junction. Then we can define the $[p,q]$-cut to run along the
geodesic $\vect{p}{q}$-prong (Fig.~\ref{fig17}(b)). 
Let the coordinate of the junction point on the $z$-plane be indicated
by $z_1$ when approached from below the cut, and by $z_2$ when
approached from above the cut ($z_1=z_2$) and  let their images under
\myref{wrs} be $w_1$ and $w_2$, respectively (see
Fig.~\ref{fig17}(c)). Our claim is that the straight distance 
between $w_1$ and $w_2$ in the $w$-plane is in fact identical to the
length of the $\vect{p}{q}$-prong, as measured with the required
$(p,q)$ metric. This is the generalized version of the result 
mentioned at the beginning of this section.   

By definition we have
\bea
w_2-w_1 &=& \int_{z_1}^{z_2}h_{\vect{r}{s}}(z)dz \,,
\label{step1}
\eea
where the path of the integration surrounds $z_0$ in a clockwise
direction. Let us choose this path so that it runs along the prong
just below the cut from $z_1$ to $z_0$ and just above it from $z_0$ to
$z_2$. Then \myref{step1} can be written as
\bea
w_2-w_1 &=& \int_{z_1}^{z_0} h_{\vect{r}{s}}^{-}(\tau(z),z)dz
-\int_{z_1}^{z_0} h_{\vect{r}{s}}^{+}(\tau(z),z)dz \,,
\label{step2}
\eea
where we parameterized both halves of the integration curve by the same
variable (that runs from $z_1$ to $z_0$), and the values of the metric
on the two sides have been distinguished by the $\pm$ superscripts.
The discontinuity of the metric can be expressed in terms of the
discontinuity of $\tau$, which is given by the 
$\mbox{SL}(2,\bbbz)$-matrix defining 
the $[p,q]$-brane
\bea
w_2-w_1 &=& \int_{z_1}^{z_0}\left(
h_{\vect{r}{s}}^{-}(\tau(z),z)
-h_{\vect{r}{s}}^{-}(M^{-1}_{p,q}\tau(z),z)\right)dz\,.
\label{step3}
\eea
{}From the $ \mbox{SL}(2,\bbbz)$-invariance of the metric ($M$ is any
$\mbox{SL}(2,\bbbz)$-matrix), 
\bea
h_{\vect{r}{s}}(\tau,z) &=& h_{M\vect{r}{s}}(M\tau,z)
\eea
it then follows that 
\bea
w_2-w_1 &=& \int_{z_1}^{z_0}\left(
h_{\vect{r}{s}}^{-}(\tau(z),z)
-h_{M_{p,q}\vect{r}{s}}^{-}(\tau(z),z)\right)dz \,.
\label{step4}
\eea
Because of \myref{chcons} and the linear dependence of the 
metric on the $p,q$ labels, we have 
\bea
h_{M_{p,q}\vect{r}{s}} = h_{\vect{r+p}{s+q}} = h_{\vect{r}{s}} +
h_{\vect{p}{q}} \eea
which leads to the final formula
\bea
w_2-w_1 &=& -\int_{z_1}^{z_0}
h_{\vect{p}{q}}(\tau(z),z)dz\,.
\label{step5}
\eea
We have thus been able to relate the $w$-coordinate (which is tied to
the $\vect{r}{s}$-metric and is the convenient arena for the
$\vect{r}{s}$-strings) to the $\vect{p}{q}$-metric which one uses to
determine the length of the $\vect{p}{q}$-prong. In particular the
length of the geodesic $\vect{p}{q}$-string between the junction
point and $z_0$ is given as
\bea
l_{\vect{p}{q}} = |\int_{z_1}^{z_0} h_{\vect{p}{q}}(\tau(z),z)dz|
= |w_2-w_1|\,,
\eea
which is the true length of the line segment between
$w_1$ and $w_2$ in the $w$-plane inside the shaded wedge. This
generalizes our previous statement about the mass of the string
junctions, and in effect guarantees that for those situations where a
string junction exists, the corresponding direct string will be longer
and will not be smooth.

\section{On the absence of smooth indirect geodesics}
\label{sec:needaprong}
\setcounter{equation}{0}

In the previous sections we showed through several examples that one
may arrange the 7-branes in the IIB moduli space such that the
would-be BPS open strings fail to be smooth geodesics and
are replaced by pronged objects. The purpose of
this section is to show explicitly that open strings can 
even fail to be smooth geodesics in the 
familiar resolution of the $so(8)$ singularity ($D_4$) which is
related to the $su(2)$ Seiberg-Witten theory with
$N_f=4$ \cite{seibergwitten}.\footnote{Our arguments  
here are related to and inspired by those of \cite{bergman}
where it is shown that in $su(2)$ SW-theory, realized as
the world-volume theory of a three brane in the vicinity of two
seven-branes, one needs a pronged object to represent the W-boson.} 
This result confirms once more that the necessity of pronged objects
is not limited to the case of exceptional groups.

Consider the resolved $so(8)$ singularity, but for the sake of
simplicity let us restrict ourselves to the case where the four 
{\bf A}-branes coincide. There are then three singular points on the
sphere (the {\bf B} brane, the {\bf C} brane, and the four {\bf A}
branes), and the moduli space is that of the $su(2)$ Seiberg-Witten  
theory with four flavors of equal masses \cite{seibergwitten}.  
Let $a(z)$ and $a_D(z)$ be the usual multivalued functions of the
sphere which transform as an SL(2,Z) doublet, $\vect{a_D}{a}$ under
the monodromies of $\tau$, where
\bea
\frac{da_D(z)}{da(z)} = \tau(z).
\eea
In the conventions of section 2.2, $\vect{a_D}{a}$ changes to 
$\vect{a_D}{a}\to K\vect{a_D}{a}$ as we change a branch cut
anti-clockwise.

The case when the {\bf B} and {\bf C} brane are on top of each other
and the {\bf A}'s are separated was studied in detail in
\cite{senorientifold,GZ}, and the BPS states were shown to be direct
and indirect strings between the {\bf A}-branes. The latter are those
which start on one of the four {\bf A}-branes and go to one of the
other {\bf A}-branes after encircling the conical singularity of the
{\bf BC} singularity. Since we are considering the limiting case where
all four {\bf A}-branes coincide, all indirect strings hit the apex,
but the strings that begin and end on different {\bf A}-branes become
smooth geodesics as we resolve the four {\bf A} branes
infinitesimally, whereas the strings that begin and end on the same
{\bf A} brane continue to go through the apex. (From an F-theory point
of view, these last geodesics do not correspond to cycles with
self-intersection number $(-2)$; they have self-intersection number  
$(-4)$ and thus do not correspond to simple spheres.)

As the {\bf B} and {\bf C} branes are resolved, we may expect (and
this was conjectured in \cite{GZ}) that the indirect {\bf A-A} strings
that begin and end on different nearby {\bf A}-branes do not necessarily
remain smooth geodesics, but that some become string junctions. In the
following we shall demonstrate that this seems indeed to be the case: 
when the two {\bf A} branes coincide, there is no smooth geodesic in
the homotopy class of the indirect string, irrespective of whether
{\bf B} and {\bf C} coincide or not.  As a consequence, when {\bf B}
and {\bf C} are finitely separated we expect the indirect string to
fail to be smooth for small separation  of the two {\bf A} branes.

\onefigure{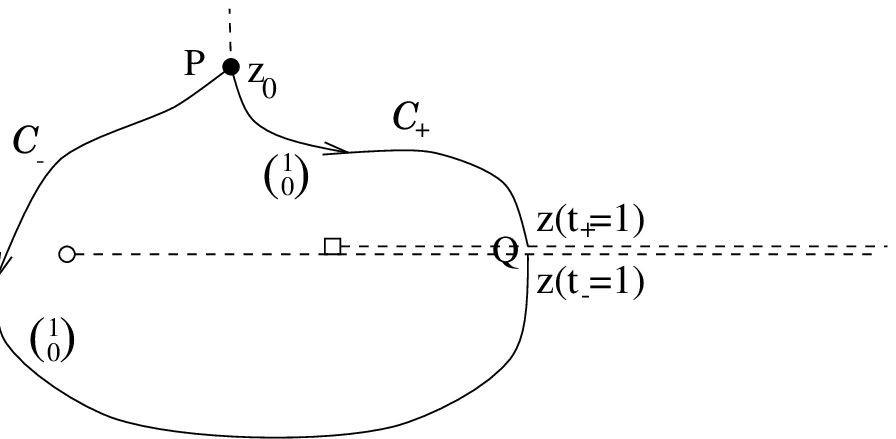}{The indirect {\bf A-A} string can be described
     as two $\vect{1}{0}$-strings departing $z_0$ which meet smoothly
     on both sides of the unified {\bf BC}-cut. \label{fig19}}
In the homotopy class of the indirect {\bf A-A} string, the 
$\vect{1}{0}$-string crosses the branch cuts of both the {\bf B} and
the {\bf C} brane before returning as a $\vect{-1}{0}$-string. To
simplify the analysis, let us choose the cuts for the {\bf B} and the
{\bf C} so that they run on top of each other where possible as in
Fig.~\ref{fig19}. The point $P$ with coordinate $z_0$ represents the
location of the {\bf A} branes, and the two oriented curves 
${\cal C}_-$ and ${\cal C}_+$ that extend from $P$ to the point $Q$
from the bottom and top of the branch cuts, respectively, each
represent part of the indirect path where the string is described as a
$({1\atop 0})$-string. In order for the complete loop 
${\cal C}_+ -{\cal C}_-$ to be a smooth geodesic, both ${\cal C}_+$
and ${\cal C}_-$ must be geodesic segments extending from $P$ to $Q$,
and the two segments must join smoothly at $Q$ (where $Q$ should not
coincide with any of the branes if the geodesic is to be smooth).

According to \cite{senthreebrane}, and taking account of our
conventions, in the coordinate
\bea
\label{coordineq}
dw_{p,q} &=& pda - qda_D \,,  
\eea
$\vect{p}{q}$ geodesic strings are straight lines. The curves 
${\cal C_+}$ and ${\cal C_-}$, both being $\vect{1}{0}$-strings, are
therefore straight lines in the coordinate $w$, where $w$ is defined
as    
\bea
\label{wtoz}
dw(z) &=& da(z)\,. 
\eea
Let us now introduce parameters $t_+, t_- \in [0,1]$ for the 
lines $w_+ (t_+)$  and $w_-(t_-)$ describing ${\cal C_+}$ and 
${\cal C_-}$, respectively, and write 
\bea
\label{pair}
w_+(t_+) &=& w_0+ t_+(w_+(1)-w_0) \,,\\
w_-(t_-) &=& w_0+ t_-(w_-(1)-w_0)  \,,
\eea
where $w_0=w(z_0)$ is the starting point of both segments. As the
monodromy around the {\bf CB}-system is  
\bea
K_C K_B = - T^4 = \mat{-1}{-4}{0}{-1} \,, 
\eea
the values of $a$ above and below the cut are
related to each other as  $a_+(z) = -a_-(z)$. The condition that the
two segments join smoothly at $Q$ requires then that
\bea
\label{scond}
\left(\frac{dz}{dt}\right)_+(Q) = 
- \alpha_0 \left(\frac{dz}{dt}\right)_- (Q)\,,
\eea
where the minus sign is due to the parameterizations used, and
$\alpha_0$ is a real positive constant. On the other hand
using \myref{wtoz} and \myref{pair} we find
\bea
\left(\frac{dz}{dt}\right)_+ 
= 
\frac{\left(\frac{dw}{dt}\right)_+}{\left(\frac{da}{dz}\right)_+}
=
\frac{w_+(1)-w_0}{-\left(\frac{da}{dz}\right)_-}
=
- \frac{w_+(1)-w_0}{w_-(1)-w_0}\left(\frac{dz}{dt}\right)_- \, .
\eea
Comparing with \myref{scond} this implies that 
\bea
\hbox{Arg} \,(w_+(1)-w_0) = \hbox{Arg}\, (w_-(1)-w_0) \,,
\eea
and the two curves ${\cal C}_+$ and ${\cal C}_-$ must have
exactly the same slope at their coincident origin $z_0$.
A smooth geodesic, however, is uniquely determined by a initial point
and its slope at this point, and thus the ${\cal C}_+$ and 
${\cal C}_-$ curves in the $z$ plane coincide, unless they encounter a
singularity. If the geodesics are to be smooth, they do not encounter
any singularity, and it follows that these indirect geodesics cannot
be smooth.

\section{Conclusions and open questions}

In this paper we have examined type IIB superstring compactified on a
two-sphere in the background of parallel 7-branes. We have shown that
the open string geodesics which represent BPS states in a given
background arrangement of 7-branes may fail to be smooth as the
background is changed. The corresponding state then does not disappear
from the spectrum, but instead it is represented by a different
object, a geodesic three-string junction or another open string.
This property guarantees that the number of BPS states that
correspond to gauge vectors in the limit of the enhanced symmetry is
independent of the position in moduli space. 
It should also be emphasized that these multipronged
objects are in general necessary to describe the BPS spectrum, and
this phenomenon is relevant not only to exceptional gauge groups.

The mechanism by means of which the different geodesic objects
transform into one another is most easily understood in the subspace
of the moduli space where the modular parameter $\tau$ is constant. It
is then possible to analyze the transitions by means of elementary
geometry, and the existence and uniqueness of the representative is 
manifest. This gives an explicit and concrete illustration of (one of
the versions of) the Hanany-Witten effect. 

We also demonstrated that all essential features survive in the
general case where $\tau$ is not constant and the system is much less
tractable. It would be interesting to analyze this case further: in
particular, the discontinuities in the BPS spectrum of states in four
dimensional supersymmetric gauge theories that are described using
3-brane probes in the backgrounds of 7-branes
\cite{senthreebrane,bergman} may be understood using similar
techniques. It would also be of interest to try to understand the
spectrum of massive BPS states. Such states should presumably be
represented by geodesics that remain of finite length as the branes
collapse to form the singularities with enhanced gauge symmetry.
\smallskip

At a more conceptual level, our work suggests that in type IIB
superstring theory, string junctions are not only necessary
ingredients for the description of the theory, but are, in fact, 
on the same footing as the $({p \atop q})$ strings, as
there exist transitions between them. A good understanding of the
physics of the junctions may therefore guide the way to a proper
formulation of type IIB superstring theory.

\section*{Acknowledgments}

\noindent We wish to thank O.~Bergman and A.~Fayazzudin for discussion
of their research prior to publication. We are also happy to
acknowledge useful conversations with O.~Aharony, M.~Berkooz,
O.~DeWolfe, J.~Goldstone, M.B.~Green, A.~Hanany, A.~Iqbal and
C.~Vafa.

\smallskip

\noindent M.R.G. is supported by a Research Fellowship of Jesus
College, Cambridge. T.H. and B.Z. are supported in part by
D.O.E. contract DE-FC02-94ER40818.

\end{document}